\documentclass[preprint,superscriptaddress,nofootinbib]{revtex4}

  \usepackage{amssymb,amsmath,graphicx,subfigure,color,rotating}
  \usepackage[colorlinks]{hyperref}
  \usepackage[left]{lineno}
  \allowdisplaybreaks[4]
  \begin{document}

  \title{Reinvestigating the $B$ ${\to}$ $PP$ decays by including
  the contributions from ${\phi}_{B2}$}
  \author{Yueling Yang}
  \affiliation{Institute of Particle and Nuclear Physics,
              Henan Normal University, Xinxiang 453007, China}
  \author{Lan Lang}
  \affiliation{Institute of Particle and Nuclear Physics,
              Henan Normal University, Xinxiang 453007, China}
  \author{Xule Zhao}
  \affiliation{Institute of Particle and Nuclear Physics,
              Henan Normal University, Xinxiang 453007, China}
  \author{Jinshu Huang}
  \affiliation{School of Physics and Electronic Engineering,
              Nanyang Normal University, Nanyang 473061, China}
  \author{Junfeng Sun}
  \affiliation{Institute of Particle and Nuclear Physics,
              Henan Normal University, Xinxiang 453007, China}

  \begin{abstract}
  Considering the $B$ mesonic distribution amplitude ${\phi}_{B2}$,
  we reinvestigated the $B$ ${\to}$ $PP$ (where $P$ $=$ ${\pi}$
  and $K$) decays with the perturbative QCD (pQCD) approach
  based on the $k_{T}$ factorization for three scenarios.
  It is found that the contributions of ${\phi}_{B2}$
  to formfactors $F_{0}^{B{\to}P}(0)$ and branching ratios are
  comparable with those from the NLO corrections.
  The $B$ ${\to}$ $K{\pi}$ decays could be well explained
  by considering the ${\phi}_{B2}$.
  Hence, when the nonleptonic $B$ decays are studied with
  the pQCD approach, the ${\phi}_{B2}$ should be taken into
  account seriously.

  \href{https://doi.org/10.1103/PhysRevD.103.056006}
       {https://doi.org/10.1103/PhysRevD.103.056006}
  \end{abstract}
  \maketitle

  It is well known that many breakthrough discoveries have come
  from precise experiments.
  $B$ physics is on the bleeding edge and one of hot topics of
  current particle physics, because of the renewed impetus from
  the successive CLEO, BaBar, Belle, LHCb and Belle-II experiments.
  Various $B$ meson decay modes with branching ratio larger than
  $10^{-6}$ have been extensively studied by the BaBar and Belle
  Collaborations with $0.56$ $ab^{-1}$ and $1.02$ $ab^{-1}$ data
  samples in the past years \cite{pdg2020,hflav}.
  A few phenomena of inconsistencies between experimental
  measurements and theoretical expectations from the standard
  model (SM) are emerging.
  More and more $B$ meson data are expected in the near future,
  about $50$ $ab^{-1}$ by the Belle-II detector at the $e^{+}e^{-}$
  SuperKEKB collider \cite{1808.10567} and about $300$ $fb^{-1}$
  by the LHCb detector at the High Luminosity LHC (HL-LHC) hadron
  collider \cite{1808.08865}.
  Besides some new phenomena, the much more precise measurements
  of $B$ meson weak decays will offer a much more rigorous test on SM.
  When looking for a smoking gun of new physics and settling the
  temporary differences between experimental and theoretical results,
  a more careful calculation on $B$ meson decays within SM is
  very necessary and important.
  In this paper, we will reinvestigate the $B$ ${\to}$ $PP$ decays
  (here $P$ $=$ ${\pi}$ and $K$) based on the perturbative QCD
  approach within SM, by considering the contributions from $B$
  mesonic wave function ${\phi}_{B2}$ which usually attract
  less attention in previous calculation.

  For clarity, we will sketch the phenomenological study
  of nonleptonic $B$ ${\to}$ $PP$ decays, although
  they have been extensively studied, for example, in Refs.
  \cite{prd65.074001,npb675.333,prd72.074007,prd80.114008,
  jpg38.015006,plb504.6,prd72.114005,prd90.074018,prd74.034010}.
  Because of our inadequate comprehension of the flavor mixing
  and possible glueball components, the final states of
  ${\eta}$ and ${\eta}^{\prime}$ mesons are not considered
  here for the moment.

  At the quark level, based on the operator product expansion and
  renormalization group (RG) method, the effective Hamiltonian
  responsible for $B$ ${\to}$ $PP$ decays is written as
  \cite{rmp68.1125},
   \begin{equation}
   \mathcal{H}_{\rm eff}\, =\,
   \frac{G_{F}}{\sqrt{2}}\, \sum\limits_{q=d,s}
   \Big\{ V_{ub}\,V_{uq}^{\ast} \sum\limits_{i=1}^{2}C_{i}\,Q_{i}
   - V_{tb}\,V_{tq}^{\ast} \sum\limits_{j=3}^{10}C_{j}\,O_{j} \Big\}
   +{\rm h.c.}
   \label{hamilton},
   \end{equation}
  where $G_{F}$ ${\simeq}$ $1.166{\times}10^{-5}\,{\rm GeV}^{-2}$
  \cite{pdg2020} is the Fermi weak coupling constant.
  With the Wolfenstein parametrization, the related
  Cabibbo-Kobayashi-Maskawa (CKM) factors are written
  as follows.
    \begin{eqnarray}
    V_{ub}\,V_{ud}^{\ast} &=&
    A\,{\lambda}^{3}\,({\rho}-i\,{\eta})\,
    (1-\frac{1}{2}\,{\lambda}^{2}-\frac{1}{8}\,{\lambda}^{4})
    +\mathcal{O}({\lambda}^{8})
    \label{ckm-vub-vud}, \\
    V_{tb}\,V_{td}^{\ast} &=&
    A\,{\lambda}^{3}
    +A^{3}\,{\lambda}^{7}\,({\rho}-i\,{\eta}-\frac{1}{2})
    -V_{ub}\,V_{ud}^{\ast}
    +\mathcal{O}({\lambda}^{8})
    \label{ckm-vtb-vtd}, \\
    V_{ub}\,V_{us}^{\ast} &=&
    A\,{\lambda}^{4}\,({\rho}-i\,{\eta})
    +\mathcal{O}({\lambda}^{8})
    \label{ckm-vub-vus}, \\
    V_{tb}\,V_{ts}^{\ast} &=&
    -A\,{\lambda}^{2}\,(1-\frac{1}{2}\,{\lambda}^{2}
    -\frac{1}{8}\,{\lambda}^{4})
    +\frac{1}{2}\,A^{3}\,{\lambda}^{6}
    -V_{ub}\,V_{us}^{\ast}
    +\mathcal{O}({\lambda}^{8})
    \label{ckm-vtb-vts},
    \end{eqnarray}
  and the latest values of the four Wolfenstein parameters
  ($A$, ${\lambda}$, ${\rho}$ and ${\eta}$) from data with the
  CKMfitter method \cite{pdg2020} are listed in Table \ref{input-parameter}.
  The Wilson coefficients, $C_{i}$, are perturbatively calculable
  at the scale of ${\cal O}(m_{W})$ and then evolved to the $b$
  quark decay scale ${\cal O}(m_{b})$ with the RG equation
  \cite{rmp68.1125}.
  The combinations of the well determined $G_{F}$, CKM factors
  and $C_{i}$ could be regarded as the universal and effective
  couplings of the operators $O_{i}$. The tree operators $O_{1,2}$,
  QCD penguin operators $O_{3{\sim}6}$ and electromagnetic penguin
  operators $O_{7{\sim}10}$ are local four-quark interactions
  and expressed as follows.
   \begin{eqnarray}
   O_{1} &=&
   (\bar{u}_{\alpha} b_{\alpha})_{V-A}
   (\bar{q}_{\beta}  u_{\beta})_{V-A}
   \label{operator-01}, \\
   O_{2} &=&
   (\bar{u}_{\alpha} b_{\beta})_{V-A}
   (\bar{q}_{\beta} u_{\alpha})_{V-A}
   \label{operator-02}, \\
   O_{3} &=&
    (\bar{q}_{\alpha} b_{\alpha})_{V-A} \sum\limits_{q^{\prime}}
    (\bar{q}^{\prime}_{\beta} q^{\prime}_{\beta})_{V-A}
   \label{operator-03}, \\
   O_{4} &=&
    (\bar{q}_{\alpha}b_{\beta})_{V-A} \sum\limits_{q^{\prime}}
    (\bar{q}^{\prime}_{\beta}q^{\prime}_{\alpha})_{V-A}
   \label{operator-04}, \\
   O_{5} &=&
    (\bar{q}_{\alpha}b_{\alpha})_{V-A} \sum\limits_{q^{\prime}}
    (\bar{q}^{\prime}_{\beta}q^{\prime}_{\beta})_{V+A}
   \label{operator-05}, \\
   O_{6} &=&
    (\bar{q}_{\alpha}b_{\beta})_{V-A} \sum\limits_{q^{\prime}}
    (\bar{q}^{\prime}_{\beta}q^{\prime}_{\alpha})_{V+A}
   \label{operator-06}, \\
   O_{7} &=&
    (\bar{q}_{\alpha}b_{\alpha})_{V-A}
     \sum\limits_{q^{\prime}} \frac{3}{2}Q_{q^{\prime}}
    (\bar{q}^{\prime}_{\beta}q^{\prime}_{\beta})_{V+A}
   \label{operator-07}, \\
   O_{8} &=&
    (\bar{q}_{\alpha}b_{\beta})_{V-A}
     \sum\limits_{q^{\prime}} \frac{3}{2}Q_{q^{\prime}}
    (\bar{q}^{\prime}_{\beta}q^{\prime}_{\alpha})_{V+A}
   \label{operator-08}, \\
   O_{9} &=&
    (\bar{q}_{\alpha}b_{\alpha})_{V-A}
     \sum\limits_{q^{\prime}} \frac{3}{2}Q_{q^{\prime}}
    (\bar{q}^{\prime}_{\beta}q^{\prime}_{\beta})_{V-A}
   \label{operator-09}, \\
   O_{10} &=&
    (\bar{q}_{\alpha}b_{\beta})_{V-A}
     \sum\limits_{q^{\prime}} \frac{3}{2}Q_{q^{\prime}}
    (\bar{q}^{\prime}_{\beta}q^{\prime}_{\alpha})_{V-A}
   \label{operator-10},
   \end{eqnarray}
  where $(\bar{q}_{1}q_{2})_{V{\pm}A}$ ${\equiv}$
  $\bar{q}_{1}{\gamma}^{\mu}(1{\pm}{\gamma}_{5})q_{2}$;
  ${\alpha}$ and ${\beta}$ is the color indices;
  $Q_{q^{\prime}}$ is the electric charge of quark
  $q^{\prime}$ in the unit of ${\vert}e{\vert}$;
  and $q^{\prime}$ ${\in}$ \{$u$, $d$, $c$, $s$, $b$\}.

  The hadronic matrix elements (HMEs), ${\langle}O_{i}{\rangle}$
  ${\equiv}$ ${\langle}P_{1}P_{2}{\vert}O_{i}{\vert}B{\rangle}$,
  describe the transformations from the quarks to
  hadrons. The calculation of HMEs is on the one hand very
  complicated due to the entanglements between perturbative and
  nonperturbative contributions, and on the other hand very sensitive
  to phenomenological models because of our limited knowledge of
  dynamics of hadronization and final state interactions.
  One of the main challenges is to calculate HMEs as properly as
  possible. Theoretically, the radiative corrections to HMEs should
  be appropriately included so that the strong phase angles closely
  related to $CP$ violation could be obtained.
  For nonleptonic $B$ decays,
  the HMEs are usually written as the product of the rescattering
  amplitudes of quarks (which are calculable order by order with
  perturbation theory in principle) and wave functions of
  participating hadrons (where nonperturbative contributions are
  housed) with the fashionable QCD-inspired phenomenological models,
  either the QCD factorization (QCDF) approach \cite{prl83.1914,
  npb591.313,npb606.245,plb488.46,plb509.263,prd64.014036}
  based on the collinear approximation or the perturbative QCD (pQCD)
  approach \cite{prl74.4388,plb348.597,prd52.3958,prd63.074006,
  prd63.054008,prd63.074009,plb555.197}
  retaining the effects of transverse momentum $k_{T}$.
  Hadronic wave functions (WFs) or distribution amplitudes (DAs)
  are independent of specific process and determined from data,
  which enable evaluating HMEs to simplify greatly.

  WFs and/or DAs are the essential ingredients of the master
  formulas for evaluating HMEs with the QCDF and pQCD approaches.
  The $B$ mesonic WFs are generally composed of two scalar
  functions \cite{prd55.272,npb592.3} and written as follows
  with the convention of Refs. \cite{epjc28.515,prd74.014027}.
   \begin{eqnarray} & &
  {\langle}\,0\,{\vert}\,\bar{q}_{\alpha}(z)\,b_{\beta}(0)\,
  {\vert}\,\overline{B}(p)\,{\rangle}
   \nonumber \\ &=&
   +\frac{i}{4}\,f_{B}\,{\int}d^{4}k\,e^{-i\,k{\cdot}z}\,
    \Big\{ \big(\!\!\not{p}+m_{B}\big)\, {\gamma}_{5}\,
   \Big[\,\frac{\not{n}_{+}}{\sqrt{2}}\,{\phi}_{B}^{+}
  +\frac{\not{n}_{-}}{\sqrt{2}}\,{\phi}_{B}^{-} \Big] \Big\}_{{\beta}{\alpha}}
   \nonumber \\ &=&
   -\frac{i}{4}\,f_{B}\,{\int}d^{4}k\,e^{-i\,k{\cdot}z}\,
    \Big\{ \big(\!\!\not{p}+m_{B}\big)\,
   {\gamma}_{5}\, \Big[{\phi}_{B}^{+}+\frac{\not{n}_{-}}{\sqrt{2}}\,\big(
   {\phi}_{B}^{+}-{\phi}_{B}^{-} \big) \Big] \Big\}_{{\beta}{\alpha}}
    \nonumber \\ &=&
   -\frac{i}{4}\,f_{B}\,{\int}d^{4}k\,e^{-i\,k{\cdot}z}\,
    \Big\{ \big(\!\!\not{p}+m_{B}\big)\,
   {\gamma}_{5}\, \Big({\phi}_{B1}+\frac{\not{n}_{-}}{\sqrt{2}}\,
   {\phi}_{B2} \Big) \Big\}_{{\beta}{\alpha}}
   \label{wf-b-meson-01},
   \end{eqnarray}
   where the coordinate of the light quark is on the light cone
   i.e., $z^{2}$ $=$ $0$ and $z_{+}$ $=$ $0$.
   $n_{+}^{\mu}$ $=$ $(1,0,0)$ and $n_{-}^{\mu}$ $=$ $(0,1,0)$
   are the light cone vectors. $f_{B}$ is the decay constant.
   ${\phi}_{B}^{+}$ and ${\phi}_{B}^{-}$ are respectively the
   leading- and sub-leading-twist WFs.
   The properties and relations of ${\phi}_{B}^{\pm}$ are
   listed as follows.
    \begin{equation}
   {\int}_{0}^{1}\,dx\,{\phi}_{B}^{\pm}(x)\, =\, 1
    \label{wf-b-meson-02},
    \end{equation}
    \begin{equation}
   {\phi}_{B1}\, =\, {\phi}_{B}^{+}
    \label{wf-b-meson-03},
    \end{equation}
    \begin{equation}
   {\phi}_{B2}\, =\, {\phi}_{B}^{+}-{\phi}_{B}^{-}
    \label{wf-b-meson-04},
    \end{equation}
    \begin{equation}
   {\phi}_{B}^{+}(x) + x\, {\phi}_{B}^{-{\prime}}(x)\, =\, 0
    \label{wf-b-meson-05},
    \end{equation}
   where $x$ is the longitudinal momentum fraction carried
   by the light quark in the $B$ meson.
   ${\phi}_{B}^{+}$ and ${\phi}_{B}^{-}$ have different asymptotic
   behaviors as $x$ ${\to}$ $0$, ${\phi}_{B}^{+}$ ${\sim}$ $x$
   but ${\phi}_{B}^{-}$ will not vanish. So they do not coincide,
   i.e., ${\phi}_{B}^{+}$ ${\neq}$ ${\phi}_{B}^{-}$ or
   ${\phi}_{B2}$ ${\neq}$ $0$.
   In many actual calculations of nonleptonic $B$ decays, only the
   contributions of ${\phi}_{B1}$ are considered appropriately,
   while those of ${\phi}_{B2}$ are assumed to be power
   suppressed and almost completely neglected.
   However, studies of Refs. \cite{npb625.239,epjc28.515,prd74.014027,
   npb642.263,prd71.034018} have shown that contributions
   of ${\phi}_{B2}$ to the $B$ ${\to}$ ${\pi}$ transition formfactors
   with the pQCD approach could have a large proportion rather than
   negligible. For example, the share could reach up to ${\sim}$
   $30\%$ for some specific cases \cite{epjc28.515,prd74.014027}.
   Clearly, the contributions of ${\phi}_{B2}$ will have some
   impacts on branching ratios of $B$ meson decays.
   We should pay due attention to contributions of ${\phi}_{B2}$
   in pace with the improvements of measurement precision, which
   is one main motivation of this work.
   The contributions of ${\phi}_{B2}$ to the $B$ ${\to}$ $PP$ decays
   have been studied with the QCDF approach \cite{epjc.79.996}.
   The study of Ref. \cite{epjc.79.996} showed that ${\phi}_{B2}$
   only contributed to nonfactorizable annihilation amplitudes,
   and is helpful in explaining pure annihilation $B$ decays.
   Different from the QCDF case, ${\phi}_{B2}$ will contribute
   to both factorizable and nonfactorizable emission amplitudes
   with the pQCD approach, besides the nonfactorizable
   annihilation amplitudes.
   That is to say, ${\phi}_{B2}$ would have much more influence
   on nonleptonic $B$ decays with the pQCD approach when
   compared with the QCDF approach.
   However, the contributions of ${\phi}_{B2}$ to the $B$ ${\to}$
   $PP$ decays have not been studied with the pQCD approach, which
   is the focus of this paper.

   One candidate of the most often used leading $B$ mesonic WF
   ${\phi}_{B}^{+}$ in earlier studies with the pQCD approach
   is written as \cite{prd63.054008}
    \begin{equation}
   {\phi}_{B}^{+}(x,b)\, =\, N\, x^{2}\,\bar{x}^{2}\, {\exp}\Big\{
   -\Big( \frac{x\,m_{B}}{\sqrt{2}\,{\omega}_{B}} \Big)^{2}
   -\frac{1}{2} {\omega}_{B}^{2}\,b^{2} \Big\}
    \label{wf-b-meson-06},
    \end{equation}
   where $b$ is the conjugate variable of the transverse momentum
   $k_{T}$. $\bar{x}$ $=$ $1$ $-$ $x$.  ${\omega}_{B}$ is the shape
   parameter. $N$ is the normalization constant.
    \begin{equation}
   {\int}_{0}^{1}dx\, {\phi}_{B}^{+}(x,0)\, =\, 1
    \label{wf-b-meson-07}.
    \end{equation}
   The corresponding sub-leading $B$ mesonic WF ${\phi}_{B}^{-}$
   \cite{prd74.014027} can be obtained by solving the equation
   of motion given by Eq.(\ref{wf-b-meson-05}).
    \begin{eqnarray}
   {\phi}_{B}^{-}(x,b) &=& N\, \frac{2\,{\omega}_{B}^{4}}{m_{B}^{4}}\,
   {\exp}\Big(-\frac{1}{2} {\omega}_{B}^{2}\,b^{2} \Big)\,\Big\{
    \sqrt{{\pi}}\,\frac{m_{B}}{\sqrt{2}\,{\omega}_{B}}
   {\rm Erf}\Big( \frac{m_{B}}{\sqrt{2}\,{\omega}_{B}},
    \frac{x\,m_{B}}{\sqrt{2}\,{\omega}_{B}}\Big)
    \nonumber \\ & &+
    \Big[1+\Big(\frac{m_{B}\,\bar{x}}{\sqrt{2}\,{\omega}_{B}}\Big)^{2}\Big]
   {\exp}\Big[-\Big(\frac{x\,m_{B}}{\sqrt{2}\,{\omega}_{B}}\Big)^{2} \Big]
   -{\exp}\Big(-\frac{m_{B}^{2}}{2\,{\omega}_{B}^{2}} \Big) \Big\}
    \label{wf-b-meson-08}.
    \end{eqnarray}

   In addition, according to the convention of Refs.
   \cite{prd65.014007,jhep.0605.004},
   WFs of the final pseudoscalars ${\pi}^{+}$ and $K^{+}$ are
   generally written as follows.
    \begin{eqnarray} & &
   {\langle}\,M(p)\,{\vert}\, \bar{q}_{\alpha}(0)\, u_{\beta}(z)\,
   {\vert}\,0\,{\rangle}
    \nonumber \\ &=&
   -\frac{i\,f_{M}}{4}\, \Big\{ {\gamma}_{5}
    \Big[ \!\!\not{p}\,{\phi}_{M}^{a}(x)
   +{\mu}_{M}\,{\phi}_{M}^{p}(x)
   -{\mu}_{M}\,\Big( \frac{\not{p}\!\!\not{z}}{p{\cdot}z}-1\Big)\,
    {\phi}_{M}^{t}(x) \Big] \Big\}_{{\beta}{\alpha}}
    \label{wf-pi-meson-01},
    \end{eqnarray}
   where $f_{M}$ is the decay constant. $x$ is the longitudinal
   momentum fraction of the anti-quark.
   ${\mu}_{M}$ $=$ $1.6{\pm}0.2$ GeV \cite{jhep.0605.004}
   is the chiral mass.
   DA ${\phi}_{M}^{a}$ is the leading twist (twist-2), and
   ${\phi}_{M}^{p,t}$ is the twist-3.
   Their explicit expressions are given in Ref. \cite{jhep.0605.004}.
    \begin{equation}
   {\phi}_{M}^{a}(x)\, =\, 6\,x\,\bar{x}\,\big\{
    1+a_{1}^{M}\,C_{1}^{3/2}({\xi})
     +a_{2}^{M}\,C_{2}^{3/2}({\xi})\big\}
    \label{wf-pi-twsit-2},
    \end{equation}
    \begin{eqnarray}
   {\phi}_{M}^{p}(x) &=& 1+3\,{\rho}_{+}^{M}
   -9\,{\rho}_{-}^{M}\,a_{1}^{M}
   +18\,{\rho}_{+}^{M}\,a_{2}^{M}
    \nonumber \\ &+&
    \frac{3}{2}\,({\rho}_{+}^{M}+{\rho}_{-}^{M})\,
    (1-3\,a_{1}^{M}+6\,a_{2}^{M})\,{\ln}(x)
    \nonumber \\ &+&
    \frac{3}{2}\,({\rho}_{+}^{M}-{\rho}_{-}^{M})\,
    (1+3\,a_{1}^{M}+6\,a_{2}^{M})\,{\ln}(\bar{x})
    \nonumber \\ &-&
    (\frac{3}{2}\,{\rho}_{-}^{M}
    -\frac{27}{2}\,{\rho}_{+}^{M}\,a_{1}^{M}
    +27\,{\rho}_{-}^{M}\,a_{2}^{M})\,C_{1}^{1/2}(\xi)
    \nonumber \\ &+&
    ( 30\,{\eta}_{M}-3\,{\rho}_{-}^{M}\,a_{1}^{M}
    +15\,{\rho}_{+}^{M}\,a_{2}^{M})\,C_{2}^{1/2}(\xi)
    \label{wf-pi-twsit-3-p},
    \end{eqnarray}
    \begin{eqnarray}
   {\phi}_{M}^{t}(x) &=&
    \frac{3}{2}\,({\rho}_{-}^{M}-3\,{\rho}_{+}^{M}\,a_{1}^{M}
    +6\,{\rho}_{-}^{M}\,a_{2}^{M})
    \nonumber \\ &-&
    C_{1}^{1/2}(\xi)\big\{
    1+3\,{\rho}_{+}^{M}-12\,{\rho}_{-}^{M}\,a_{1}^{M}
   +24\,{\rho}_{+}^{M}\,a_{2}^{M}
    \nonumber \\ & & \quad +
    \frac{3}{2}\,({\rho}_{+}^{M}+{\rho}_{-}^{M})\,
    (1-3\,a_{1}^{M}+6\,a_{2}^{M})\,{\ln}(x)
    \nonumber \\ & & \quad +
    \frac{3}{2}\,({\rho}_{+}^{M}-{\rho}_{-}^{M})\,
    (1+3\,a_{1}^{M}+6\,a_{2}^{M})\, {\ln}(\bar{x}) \big\}
    \nonumber \\ &-&
    3\,(3\,{\rho}_{+}^{M}\,a_{1}^{M}
    -\frac{15}{2}\,{\rho}_{-}^{M}\,a_{2}^{M})\,C_{2}^{1/2}(\xi)
    \label{wf-pi-twsit-3-t},
    \end{eqnarray}
   where the variable ${\xi}$ $=$ $x$ $-$ $\bar{x}$ $=$ $2\,x$ $-$ $1$.
   The normalization conditions are
    \begin{equation}
   {\int}_{0}^{1}dx\, {\phi}_{M}^{a,p}(x)\, =\, 1
    \label{wf-pi-normalization-conditions-01},
    \end{equation}
    \begin{equation}
   {\int}_{0}^{1}dx\, {\phi}_{M}^{t}(x)\, =\,0
    \label{wf-pi-normalization-conditions-02}.
    \end{equation}
   Other parameters are expressed as \cite{{jhep.0605.004}}:
   ${\rho}_{+}^{M}$ $=$ ${\displaystyle  \frac{m_{M}^{2}}{{\mu}_{M}^{2}} }$,
   ${\rho}_{-}^{K}$ ${\simeq}$ ${\displaystyle  \frac{m_{s}}{{\mu}_{K}} }$,
   ${\rho}_{-}^{\pi}$ $=$ $0$,
   and ${\eta}_{M}$ $=$ ${\displaystyle \frac{f_{3M}}{f_{M}\,{\mu}_{M}} }$.
   The Gegenbauer polynomials are written as follows.
   \begin{equation}
   C_{1}^{1/2}({\xi})\, =\, {\xi}
   \label{gegenbauer-polynomials-1-1},
   \end{equation}
   \begin{equation}
   C_{1}^{3/2}({\xi})\, =\, 3\,{\xi}
   \label{gegenbauer-polynomials-1-3},
   \end{equation}
   \begin{equation}
   C_{2}^{1/2}({\xi})\, =\, \frac{3}{2}{\xi}^{2}-\frac{1}{2}
   \label{gegenbauer-polynomials-2-1},
   \end{equation}
   \begin{equation}
   C_{2}^{3/2}({\xi})\, =\, \frac{15}{2}{\xi}^{2}-\frac{3}{2}
   \label{gegenbauer-polynomials-2-3}.
   \end{equation}

  \begin{figure}[ht]
  \includegraphics[width=0.4\textwidth,bb=100 530 340 760]{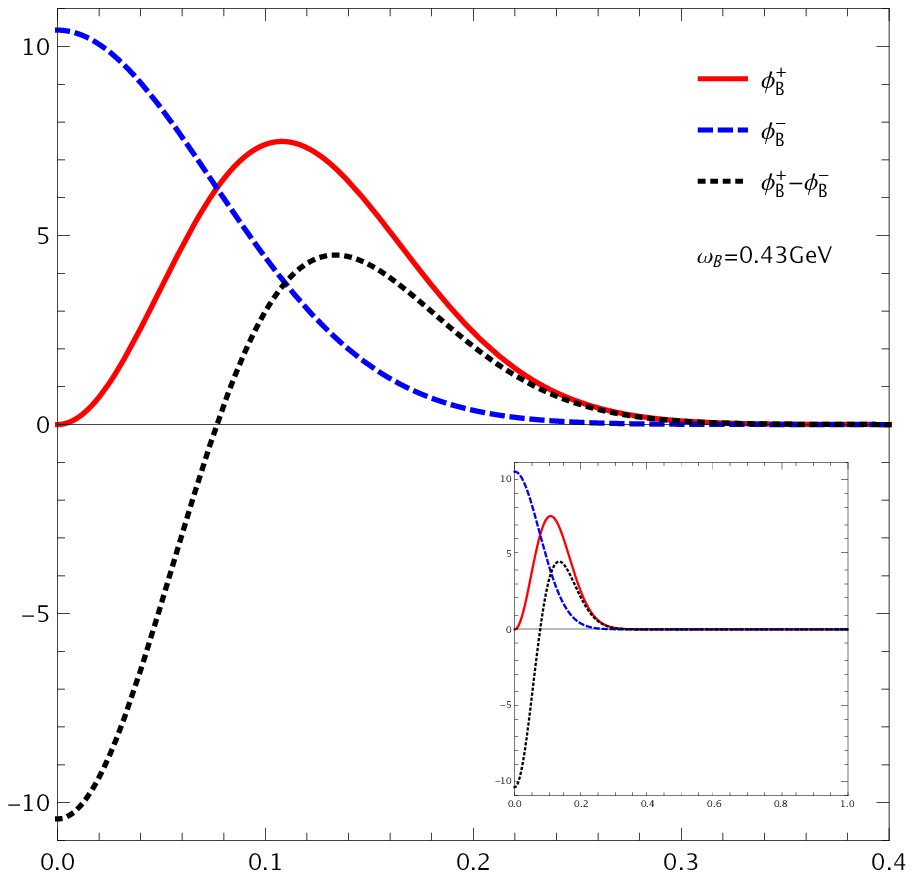}
  \caption{The shape lines of the $B$ mesonic WFs
  ${\phi}_{B}^{\pm}(x,0)$ versus $x$ (horizontal axis).}
  \label{fig-wf-b}
  \end{figure}

   The curves of the normalized DAs ${\phi}_{B}^{+}(x,0)$
   and ${\phi}_{B}^{-}(x,0)$ for $B$ meson in
   Eq.(\ref{wf-b-meson-06}) and Eq.(\ref{wf-b-meson-08})
   are displayed in Fig.\ref{fig-wf-b}.
   It can be clearly seen from Fig.\ref{fig-wf-b} that
   (1) DAs ${\phi}_{B}^{\pm}$ are very asymmetric, and peak
   at small $x$ region. This fact is generally consistent with
   the plausible suspicion that the light quark shares a
   small momentum fraction in $B$ meson. In addition,
   DAs ${\phi}_{B}^{\pm}$ vanish as $x$ ${\to}$ $1$, and thus
   offer a natural cutoff on the seemingly counterintuitive
   contributions from large $x$ domain.
   (2) ${\phi}_{B}^{-}$ and ${\phi}_{B2}$ do not vanish
   as $x$ ${\to}$ $0$, thus the integral
   ${\displaystyle {\int}dx\,\frac{{\phi}_{B2}}{x} }$ and
   ${\displaystyle {\int}dx\,\frac{{\phi}_{B2}}{x^{2}} }$
   corresponding to the factorizable emission topologies
   (form factors) diverge at the endpoint $x$ $=$ $0$,
   as discussed in Ref. \cite{npb592.3} with the collinear
   approximation. This implies that,
   on the one hand, the contributions of ${\phi}_{B2}$
   might be important at small $x$ regions and should be
   given due consideration in calculation, although
   ${\phi}_{B}^{-}$ is sub-leading twist;
   on the other hand, it seems reasonable and necessary to
   retain the contributions of the transverse momentum to regulate
   the singularities at the endpoint with the pQCD approach.

  \begin{figure}[ht]
  \includegraphics[width=0.4\textwidth,bb=100 525 340 770]{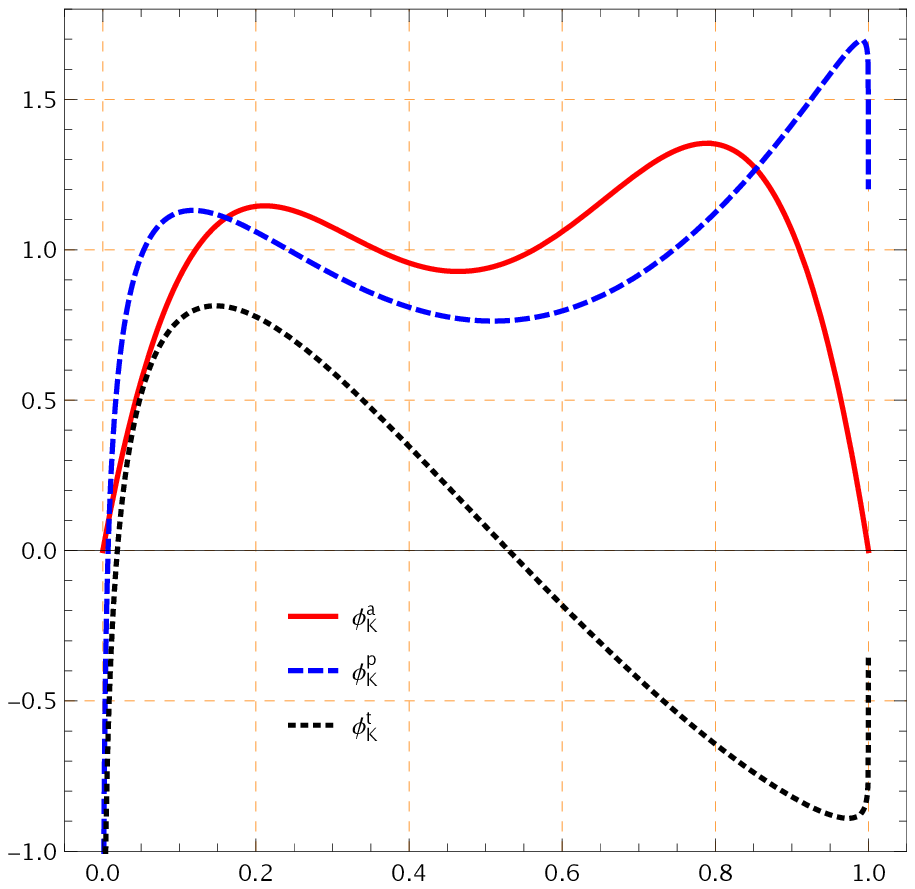}
  \qquad \quad
  \includegraphics[width=0.4\textwidth,bb=100 525 340 770]{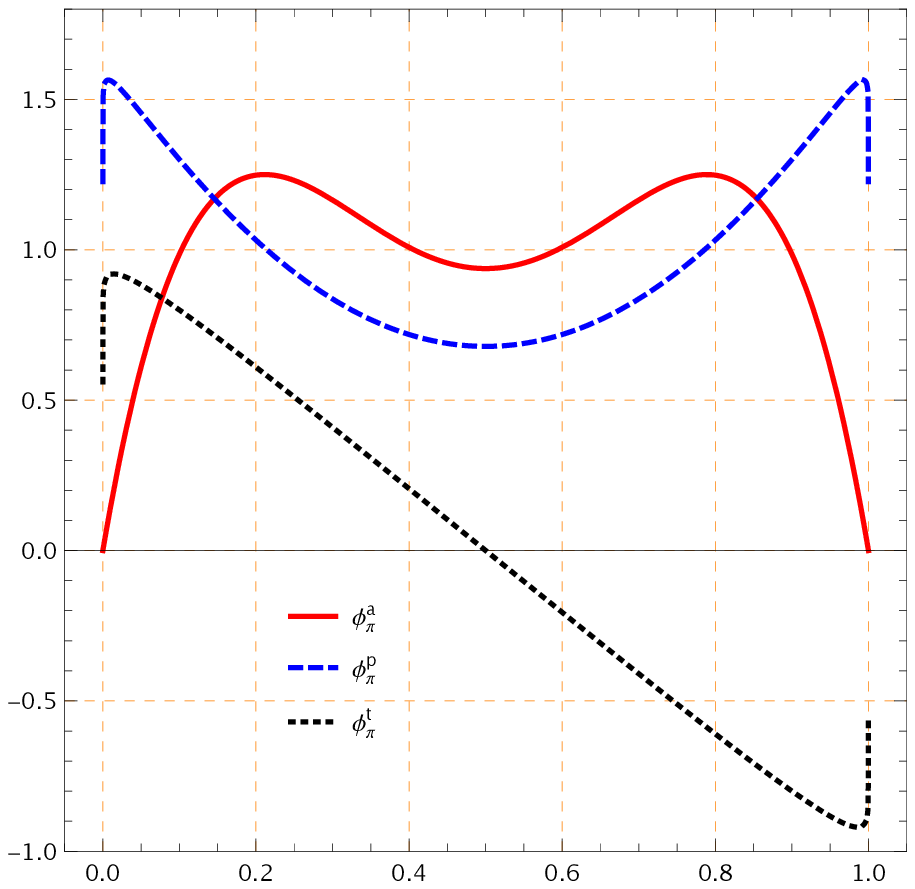}
  \\ {(a) \hspace{0.45\textwidth} (b)}
  \caption{The shape lines of DAs ${\phi}_{M}^{a,p,t}$ for $K$
   in (a) and ${\pi}$ in (b) versus $x$ (horizontal axis).}
  \label{fig-wf-pi}
  \end{figure}

   The line shapes of DAs ${\phi}_{M}^{a,p,t}(x)$ for ${\pi}$ and $K$
   mesons in Eq.(\ref{wf-pi-twsit-2}), Eq.(\ref{wf-pi-twsit-3-p})
   and Eq.(\ref{wf-pi-twsit-3-t}) are shown in Fig. \ref{fig-wf-pi}.
   The pionic DAs are totally symmetric with respect to the $x$
   ${\leftrightarrow}$ $\bar{x}$ exchange.
   The $SU(3)$ breaking effects on kaonic DAs are considered.
   The quark-mass corrections modify the asymptotic behaviors of
   ${\phi}_{M}^{p,t}$ and induce the logarithmic endpoint
   singularities, as analyzed in Ref. \cite{jhep.0605.004}.

  \begin{table}[ht]
  \caption{The values of the input parameters, where their
   central values will be regarded as the default inputs
   unless otherwise specified. The numbers in parentheses
   are errors.}
  \label{input-parameter}
  \begin{ruledtabular}
  \begin{tabular}{lll}
    CKM parameters \cite{pdg2020}
  & $A$ $=$ $0.790(17)$, 
  & ${\lambda}$ $=$ $0.22650(48)$, \\ 
  & $\bar{\rho}$ $=$ $0.141(17)$, 
  & $\bar{\eta}$ $=$ $0.357(11)$, \\ 
    mass of the particles \cite{pdg2020}
  & $m_{{\pi}^{0}}$ $=$ $134.98$ MeV,
  & $m_{{\pi}^{\pm}}$ $=$ $139.57$ MeV, \\
  & $m_{K^{0}}$ $=$ $497.61$ MeV,
  & $m_{K^{\pm}}$ $=$ $493.68$ MeV, \\
  & $m_{B_{u}}$ $=$ $5279.34(12)$ MeV, 
  & $m_{B_{d}}$ $=$ $5279.65(12)$ MeV, \\ 
  & $m_{b}$ $=$ $4.78(6)$ GeV, 
  & $m_{s}$ $=$ $130.0(1.8)$ MeV \cite{epjc.80.113}, \\ \hline
    decay constants \cite{pdg2020}
  & $f_{{\pi}}$ $=$ $130.2(1.2)$ MeV,
  & $f_{3{\pi}}$ $=$ $0.45(15){\times}10^{-2}\,{\rm GeV}^{2}$ \cite{jhep.0605.004},\\
  & $f_{K}$ $=$ $155.7(3)$ MeV,
  & $f_{3K}$ $=$ $0.45(15){\times}10^{-2}\,{\rm GeV}^{2}$ \cite{jhep.0605.004},\\
  & $f_{B}$ $=$ $190.0(1.3)$ MeV, \\ \hline
    lifetime \cite{pdg2020}
  & ${\tau}_{B^{\pm}}$ $=$ $1.638(4)$ ps,
  & ${\tau}_{B^{\pm}}$ $=$ $1.519(4)$ ps,  \\ \hline
    \multicolumn{3}{l}{Gegenbauer moments at the scale of ${\mu}$ $=$ 1 GeV \cite{jhep.0605.004}}\\
  & $a_{1}^{\pi}$ $=$ $0$,
  & $a_{2}^{\pi}$ $=$ $0.25(15)$, \\
  & $a_{1}^{K}$ $=$ $0.06(3)$,
  & $a_{2}^{K}$ $=$ $0.25(15)$. \\
  \end{tabular}
  \end{ruledtabular}
  \end{table}

   With the above mesonic DAs, we can obtain the hadron transition
   formfactors and amplitudes of the $B$ ${\to}$ $PP$ decays with
   the pQCD approach.
   There are some conventions in our calculation.
   In the rest frame of $B$ meson, the light-cone kinematic
   variables of participating particles in the heavy quark
   limit are defined as follows.
   \begin{equation}
    p_{B}\, =\, p_{1}\, =\, (p_{1}^{+},p_{1}^{-},\vec{p}_{1T})
         \, =\, \frac{m_{B}}{\sqrt{2}}(1,1,0)
   \label{kine-b-meson},
   \end{equation}
   \begin{equation}
    p_{M}\, =\, p_{2}\, =\, (p_{2}^{+},p_{2}^{-},\vec{p}_{2T})
         \, =\, \frac{m_{B}}{\sqrt{2}}(0,1,0)
   \label{kine-p2},
   \end{equation}
   \begin{equation}
    p_{M^{\prime}}\, =\, p_{3}
      \, =\, (p_{3}^{+},p_{3}^{-},\vec{p}_{3T})
      \, =\, \frac{m_{B}}{\sqrt{2}}(1,0,0)
   \label{kine-p3},
   \end{equation}
   \begin{equation}
   k_{1}\, =\, (x_{1}\,p_{1}^{+},0,\vec{k}_{1T})
   \label{kine-k1},
   \end{equation}
   \begin{equation}
   k_{2}\, =\, (0,x_{2}\,p_{2}^{-},\vec{k}_{2T})
   \label{kine-k2},
   \end{equation}
   \begin{equation}
   k_{3}\, =\, (x_{3}\,p_{3}^{+},0,\vec{k}_{3T})
   \label{kine-k3},
   \end{equation}
   where $k_{1}$ and $x_{1}$ are respectively the momentum and
   longitudinal momentum fraction of light quark in the $B$ meson;
   $k_{2,3}$ and $x_{2,3}$ are respectively the momentum and
   longitudinal momentum fraction of anti-quark in final hadrons.
   $\vec{k}_{iT}$ is the transverse momentum.
   It is clear that $p_{1}^{2}$ $=$ $m_{B}^{2}$,
   $p_{2}^{2}$ $=$ $m_{M}^{2}$ $=$ $0$ and
   $p_{3}^{2}$ $=$ $m_{M^{\prime}}^{2}$ $=$ $0$.

   The formfactors for the $B$ ${\to}$ $P$ transition
   are defined as \cite{zpc42.671}
   \begin{eqnarray} & &
  {\langle}\,M(p_{2})\,{\vert}\, (\bar{q}\, b)_{V-A}\, {\vert}
   \overline{B}(p_{1})\,{\rangle}
   \nonumber \\ &=&
   \Big\{ \big( p_{1} + p_{2} \big)^{\mu} -
   \frac{ m_{B}^{2} - m_{M}^{2} }{ q^{2} }\, q^{\mu}\, \Big\}\,
   F_{1}(q^{2})
  +\frac{ m_{B}^{2} - m_{M}^{2} }{ q^{2} }\, q^{\mu}\,
   F_{0}(q^{2})
   \label{formfactor-defination-01},
   \end{eqnarray}
   where $q$ $=$ $p_{1}$ $-$ $p_{2}$.
   It is required that $F_{0}(0)$ $=$ $F_{1}(0)$ at the
   pole of $q^{2}$ $=$ $0$.

  \begin{figure}[ht]
  \includegraphics[width=0.27\textwidth,bb=180 530 320 600]{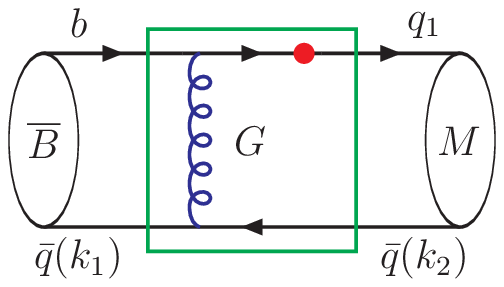}
  \qquad
  \includegraphics[width=0.27\textwidth,bb=180 530 320 600]{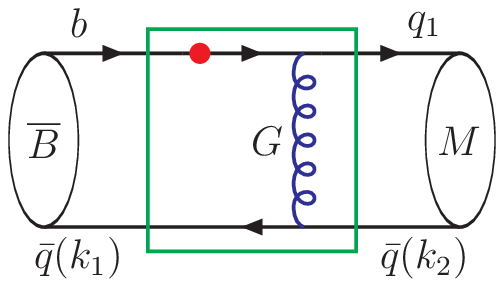}
  \\ {(a) \hspace{0.3\textwidth} (b)}
  \caption{Feynman diagrams contributing to the $\overline{B}$
   ${\to}$ $M$ formfactors, where the dots denote appropriate
   current interactions, and boxes denote quark scattering
   amplitudes.}
  \label{fey-formfactor}
  \end{figure}

\begin{figure}[ht]
  \includegraphics[width=0.4\textwidth]{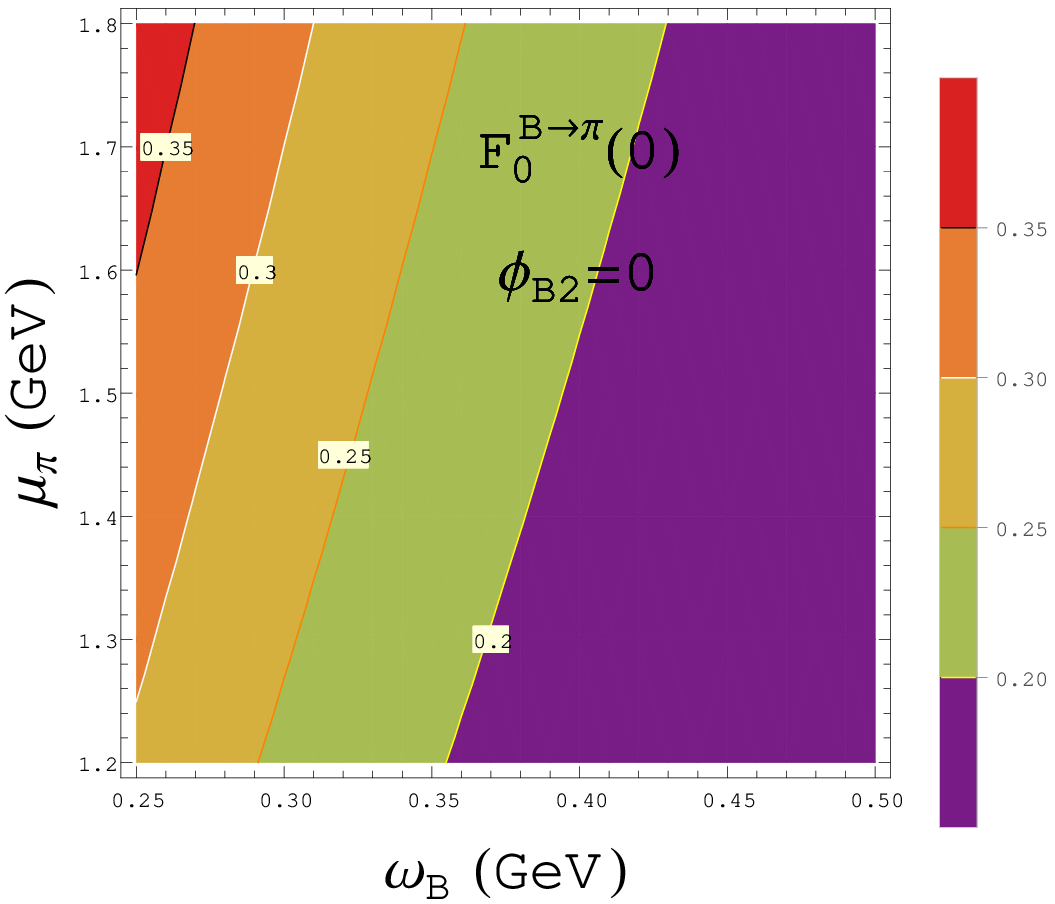}
  \quad
  \includegraphics[width=0.4\textwidth]{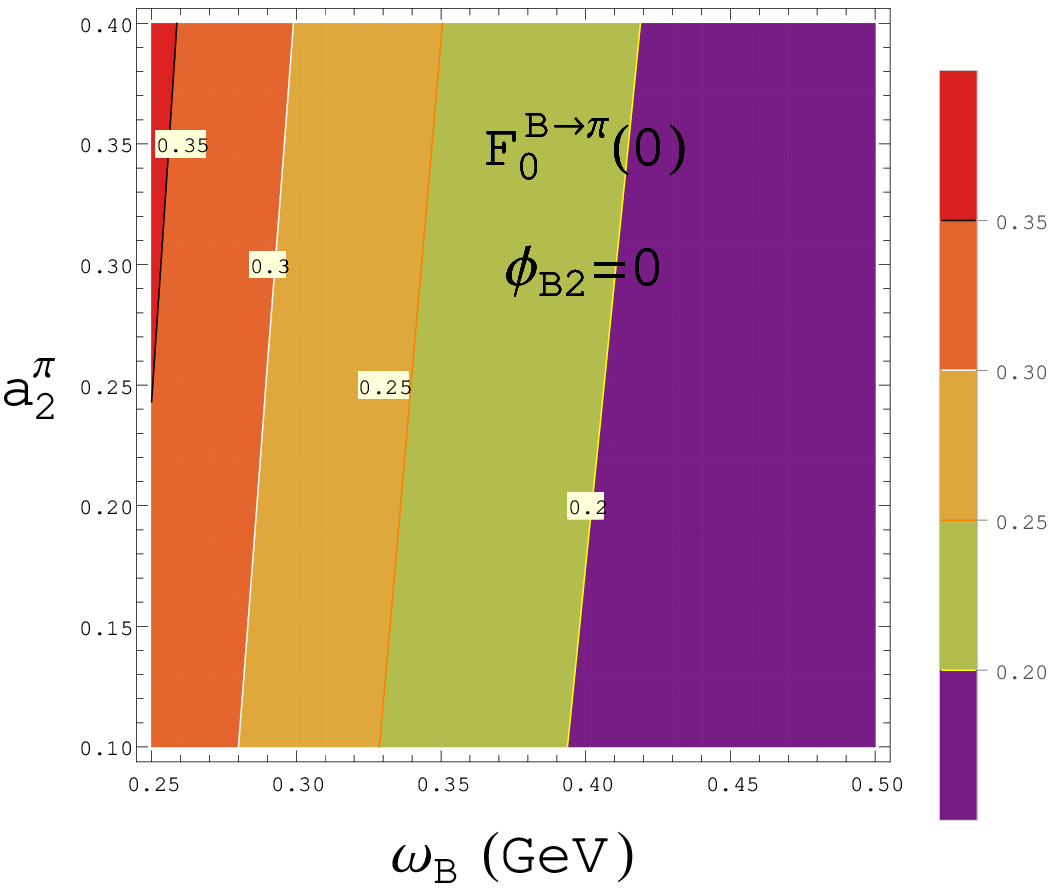}
  \\ {(a) \hspace{0.4\textwidth} (b)} \\ ~\\
  \includegraphics[width=0.4\textwidth]{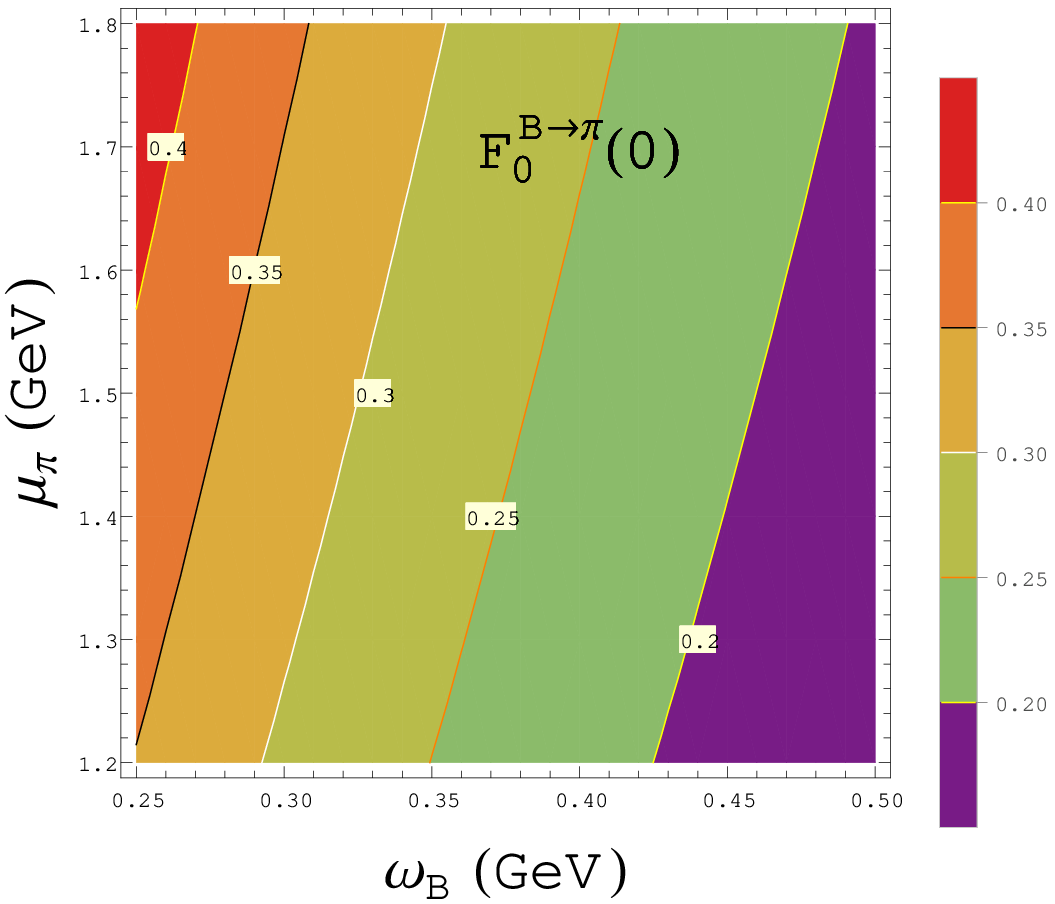}
  \quad
  \includegraphics[width=0.4\textwidth]{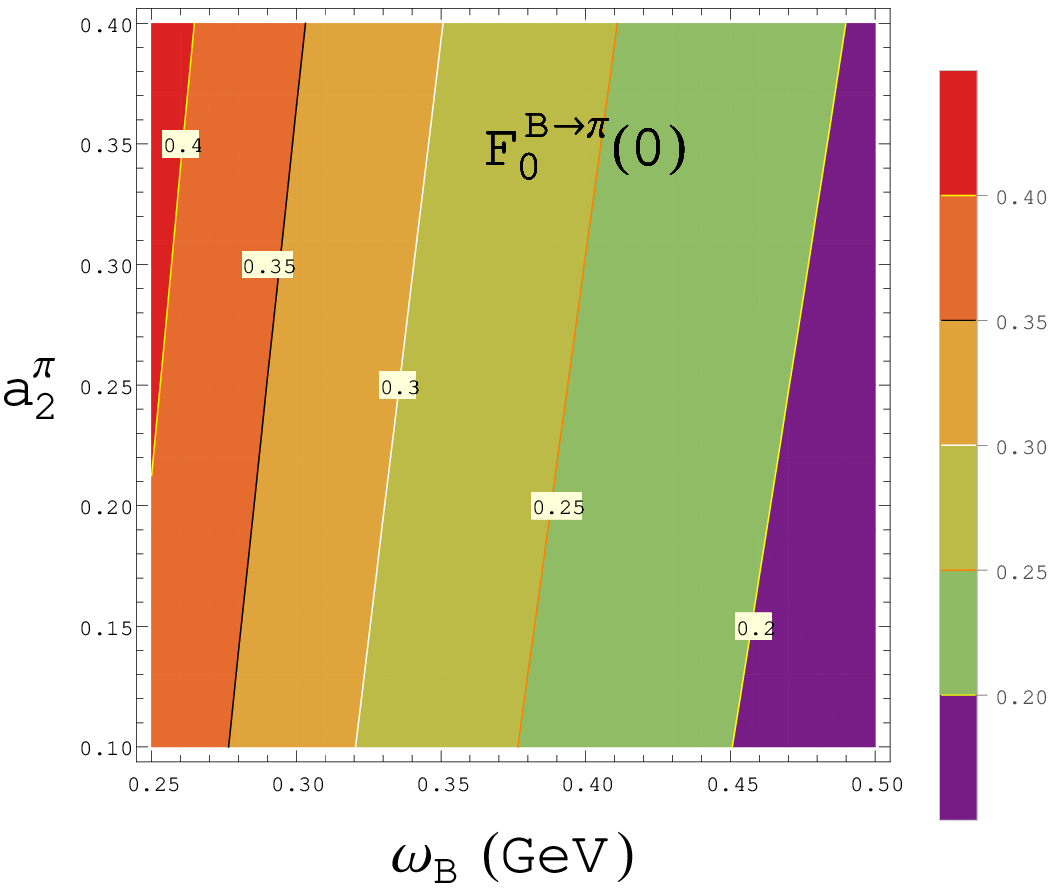}
  \\ {(c) \hspace{0.4\textwidth} (d)}
  \caption{Contour plot of $F_{0}^{B{\to}{\pi}}(0)$.
   The values of formfactors denoted by shades.
   The values in (a,b) and (c,d) are calculated without and with
   the contributions from ${\phi}_{B2}$.}
  \label{formfactor-contour}
  \end{figure}

\begin{figure}[ht]
  \includegraphics[width=0.4\textwidth]{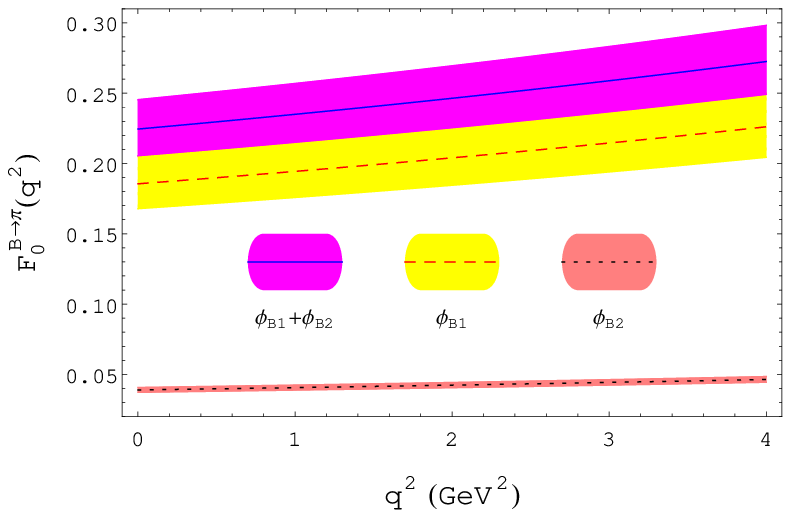}
  \quad
  \includegraphics[width=0.4\textwidth]{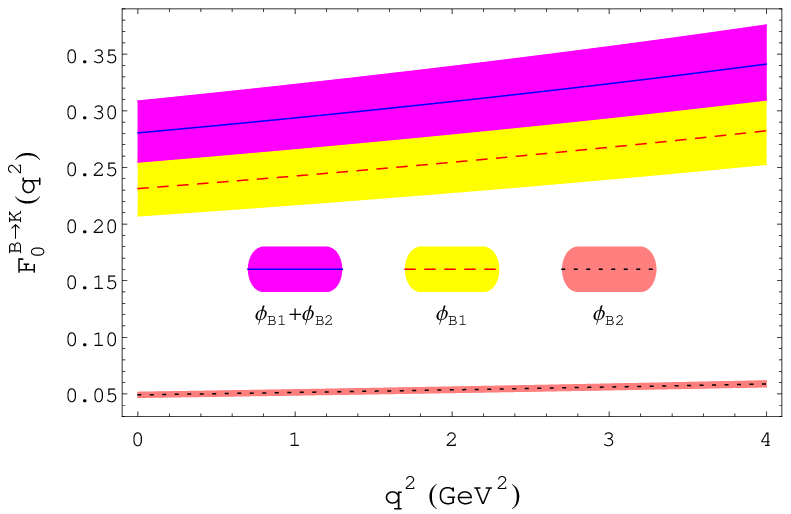}
  \\ {(a) \hspace{0.4\textwidth} (b)} \\ ~\\
  \includegraphics[width=0.4\textwidth]{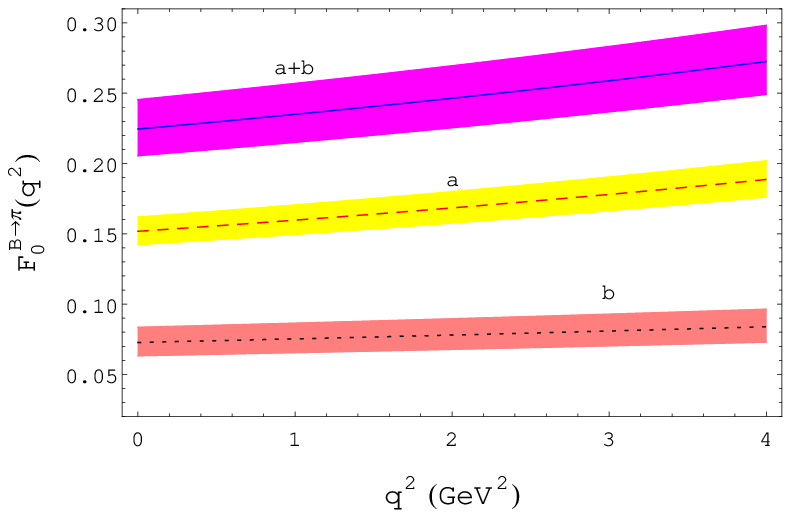}
  \quad
  \includegraphics[width=0.4\textwidth]{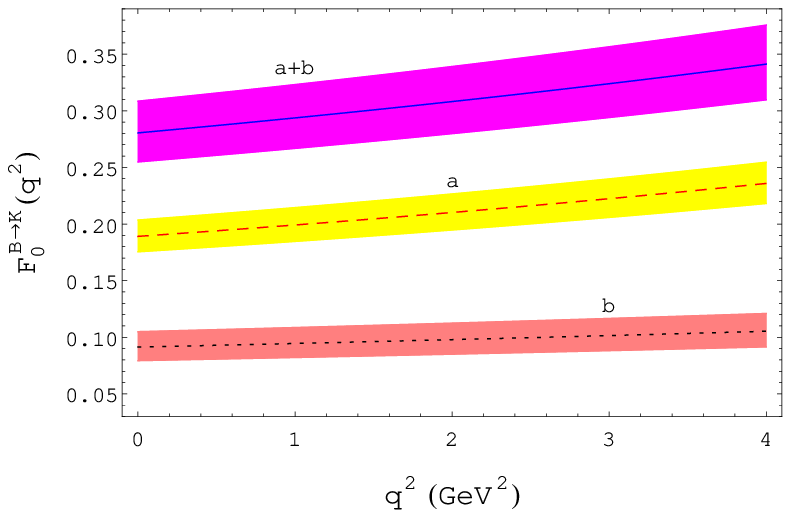}
  \\ {(c) \hspace{0.4\textwidth} (d)}
  \caption{Formfactors of $F_{0}^{B{\to}{\pi}}(q^{2})$
  and $F_{0}^{B{\to}K}(q^{2})$ versus $q^{2}$.
  The bands in (a,b) correspond to the contributions
  from WFs ${\phi}_{B1}$ and ${\phi}_{B2}$. The bands in (c,d)
  correspond to the contributions from Fig.\ref{fey-formfactor}
  (a) and (b). The bands are calculated with ${\omega}_{B}$
  $=$ $0.41$ ${\sim}$ $0.45$ GeV and ${\mu}_{M}$ $=$ $1.5$ ${\sim}$
  $1.7$ GeV. The lines correspond the S2 scenario.}
  \label{formfactor-band}
  \end{figure}

\begin{figure}[ht]
  \includegraphics[width=0.4\textwidth]{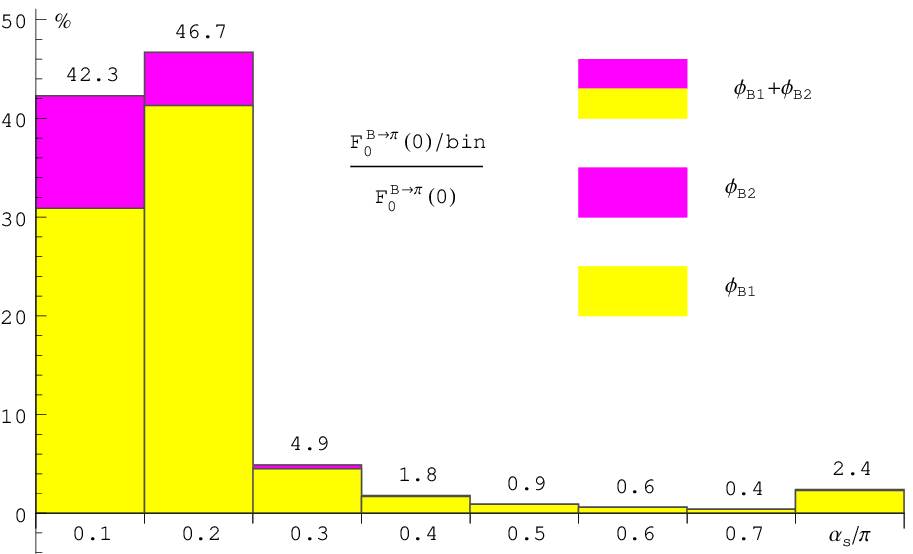}
  \quad
  \includegraphics[width=0.4\textwidth]{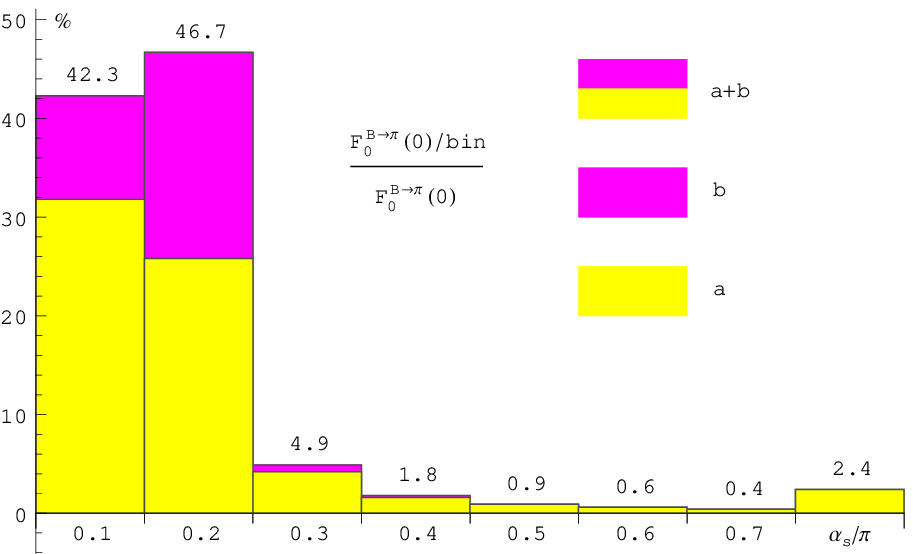}
  \\ {(a) \hspace{0.4\textwidth} (b)}
  \caption{The percentages of contributions to $F_{0}^{B{\to}{\pi}}(0)$
  from different ranges of ${\alpha}_{s}/{\pi}$ for the S2 scenario.
  The numbers over histogram denote the total percentages.
  The histograms in (a) correspond to the contributions from WFs
  ${\phi}_{B1}$ and ${\phi}_{B2}$. The histograms in (b)
  correspond to the contributions from Fig.\ref{fey-formfactor}
  (a) and (b).}
  \label{formfactor-scale}
  \end{figure}

  The lowest order Feynman diagrams for the $B$ ${\to}$ $M$
  transition formfactors are shown in Fig.\ref{fey-formfactor}.
  The formfactors $F_{i}$ are written as the convolution integrals
  of the quark scattering amplitudes $\mathcal{T}$ and hadron WFs
  ${\Phi}_{i}$ with the pQCD approach.
   \begin{equation}
   F_{i}\, =\, {\int}dx_{1}\,dx_{2}\,db_{1}\,db_{2}\,
   {\Phi}_{B}(x_{1},b_{1})\,e^{-s_{B}}\,
    \mathcal{T}(x_{1},x_{2},b_{1},b_{2})\,
   {\Phi}_{M}(x_{2},b_{2})\,e^{-s_{M}}
   \label{formfactor-defination-pqcd},
   \end{equation}
  where $b_{i}$ is the conjugate variable of transverse
  momentum $k_{iT}$.
  The Sudakov factors $e^{-s_{B}}$ and $e^{-s_{M}}$ are introduced
  for WFs ${\Phi}_{B}$ and ${\Phi}_{M}$, respectively.
  The Sudakov factor is a characteristic element and highly
  recommended by the pQCD approach to effectively regulate the
  nonperturbative contributions, so that a dominant share of
  formfactor would come from hard gluon exchange, and the
  perturbative calculation would be reasonable and practicable.
  The expressions for formfactors including the ${\phi}_{B2}$
  contributions are listed in Appendix \ref{formfactor}.
  Our results of formfactors are shown in Fig.
  \ref{formfactor-contour}, \ref{formfactor-band},
  \ref{formfactor-scale} and Table \ref{tab-formfactor}.
  \begin{table}[ht]
  \caption{The numerical values of formfactors
  $F_{0}^{B{\to}{\pi}}$ and $F_{0}^{B{\to}K}$.
  The contributions from ${\phi}_{B1}$, ${\phi}_{B2}$,
  Fig.\ref{fey-formfactor} (a) and Fig.\ref{fey-formfactor} (b)
  are given in the corresponding rows.
  The uncertainties in parentheses arise from variations of
  ${\omega}_{B}{\pm}0.01$ GeV, ${\mu}_{M}{\pm}0.1$ GeV,
  and $a_{2}^{M}{\pm}0.15$, respectively.}
  \label{tab-formfactor}
  \begin{ruledtabular}
  \begin{tabular}{c|c|ccc}
    transition & case & S1 &  S2 & S3 \\ \hline
    & ${\phi}_{B1}$
    & $0.190(06)(08)(08)$
    & $0.186(06)(07)(08)$
    & $0.181(06)(07)(07)$ \\
    & ${\phi}_{B2}$
    & $0.041(01)(00)(05)$
    & $0.039(01)(00)(04)$
    & $0.037(01)(00)(04)$  \\
      $F_{0}^{B{\to}{\pi}}(0)$
    & Fig. \ref{fey-formfactor} (a)
    & $0.156(04)(03)(13)$
    & $0.152(04)(03)(12)$
    & $0.147(04)(03)(12)$ \\
    & Fig. \ref{fey-formfactor} (b)
    & $0.074(03)(05)(00)$
    & $0.073(03)(04)(00)$
    & $0.071(03)(04)(00)$ \\ \cline{2-5}
    & total 
    & $0.231(07)(07)(13)$
    & $0.224(07)(07)(12)$
    & $0.218(06)(06)(12)$ \\ \hline
    & ${\phi}_{B1}$
    & $0.235(08)(12)(09)$
    & $0.231(07)(11)(09)$
    & $0.227(07)(10)(09)$ \\
    & ${\phi}_{B2}$
    & $0.051(01)(00)(05)$
    & $0.049(01)(00)(05)$
    & $0.047(01)(00)(05)$  \\
      $F_{0}^{B{\to}K}(0)$
    & Fig. \ref{fey-formfactor} (a)
    & $0.193(05)(06)(13)$
    & $0.189(05)(05)(14)$
    & $0.185(04)(05)(13)$  \\
    & Fig. \ref{fey-formfactor} (b)
    & $0.093(04)(06)(00)$
    & $0.091(04)(05)(01)$
    & $0.089(04)(05)(01)$  \\ \cline{2-5}
    & total 
    & $0.286(09)(12)(14)$
    & $0.280(08)(11)(14)$
    & $0.274(08)(10)(14)$
  \end{tabular}
  \end{ruledtabular}
  \end{table}

  The dependences of formfactor $F_{0}^{B{\to}{\pi}}(0)$
  on some input parameters are shown in
  Fig.\ref{formfactor-contour}.
  It is seen clearly that
  (1) formfactors obtained with the pQCD approach are sensitive
  to the shape parameter ${\omega}_{B}$ for $B$ mesonic WFs.
  This phenomenon is basically analogical with that of
  Ref. \cite{prd74.014027}.
  (2) the effects of the chiral mass ${\mu}_{M}$ indicate the
  importance of the twist-3 contributions.
  It is shown in Ref. \cite{prd74.014027} that the contributions
  from twist-3 ${\phi}_{\pi}^{p,t}$ to $F_{0}^{B{\to}{\pi}}(0)$
  could exceed 50\% with appropriate parameters.
  (3) The contributions of WF ${\phi}_{B2}$ can enhance
  the formfactors.
  By comparison of the branching ratios for $B$ ${\to}$ $PP$
  decays with the experimental results, three optimal scenarios
  are obtained, when the ${\phi}_{B2}$ is considered.

  Scenario 1 (S1) : ${\omega}_{B}$ $=$ $0.45$ GeV and
  ${\mu}_{M}$ $=$ $1.7$ GeV for PDG data;

  Scenario 2 (S2) : ${\omega}_{B}$ $=$ $0.43$ GeV and
  ${\mu}_{M}$ $=$ $1.6$ GeV for Belle data;

  Scenario 3 (S3) : ${\omega}_{B}$ $=$ $0.41$ GeV and
  ${\mu}_{M}$ $=$ $1.5$ GeV for BaBar data.

  The formfactors are assumed to be perturbatively calculable
  with the pQCD approach, but reliable only for the large
  recoil transition, i.e., the small $q^{2}$ regions.
  Thus in this paper, the formfactors with $q^{2}$
  ${\le}$ $4$ $\text{GeV}^{2}$ are calculated.
  The dependences of formfactors $F_{0}^{B{\to}{\pi}}(q^{2})$
  and $F_{0}^{B{\to}K}(q^{2})$ on $q^{2}$ and are shown in
  Fig. \ref{formfactor-band}. It is seen clearly that
  (1) the formfactors $F_{0}$ increase monotonically and
  slowly with $q^{2}$ within the large recoil domains.
  (2) The lion's share of formfactors is from B mesonic WFs
  ${\phi}_{B1}$, and the share of ${\phi}_{B2}$ is relatively
  small. This is why the contributions from B mesonic
  WFs ${\phi}_{B2}$ were usually not considered in most of
  previous works. Our results in Table \ref{tab-formfactor}
  show that the contributions from ${\phi}_{B2}$ to
  formfactors $F_{0}^{B{\to}{\pi}}(0)$ and $F_{0}^{B{\to}K}(0)$
  are about 17\%, which is much larger than 7\% from the
  next-to-leading order (NLO) contributions \cite{prd89.094004}.
  (3) More than half of the formfactors is from the contributions
  of topology Fig. \ref{fey-formfactor} (a), about 67\% shown
  in Table \ref{tab-formfactor}.

  In Fig. \ref{formfactor-scale}, more details about the
  contributions from ${\phi}_{B1}$ and ${\phi}_{B2}$,
  from topology Fig. \ref{fey-formfactor} (a) and (b)
  to formfactor $F_{0}^{B{\to}{\pi}}(0)$ at
  $q^{2}$ $=$ $0$ are displayed bin by bin
  with respect to the distributions of
  ${\displaystyle \frac{{\alpha}_{s}}{\pi} }$. It shows that
  about 90\% of formfactor comes from the region of
  ${\displaystyle \frac{{\alpha}_{s}}{\pi} }$ ${\le}$ 0.2,
  where the contributions from ${\phi}_{B1}$ and ${\phi}_{B2}$
  account for more than 70\% and 15\%, the contributions
  from Fig. \ref{fey-formfactor} (a) and (b) account for more
  than 55\% and 30\%, respectively. These results may imply
  that the quark scattering amplitudes are dominated by hard
  gluon exchange, and the perturbative calculation of the
  formfactor with the pQCD approach is feasible and reliable.
  An important and possible underlying mechanism is the way of
  choosing the hard scale as the maximum virtuality of quarks
  and gluons, see Eq.(\ref{scale-tab}), besides the
  suppression of the long-distance contributions from
  Sudakov factors.

  The values of formfactors in Table \ref{tab-formfactor} are less
  than those of Refs. \cite{prd74.014027,epjc28.515}, due to
  different DAs models and different values of input parameters.
  As is shown in Fig. \ref{formfactor-contour}, the formfactors
  decrease with the increase of shape parameter ${\omega}_{B}$.
  A large shape parameter ${\omega}_{B}$ for B mesonic WFs is used
  in our calculation, compared with that in Ref. \cite{prd74.014027}.
  It should be pointed out that a relatively small value of formfactor
  $F_{0}^{B{\to}{\pi}}(0)$ has recently been obtained by fitting
  the Bourrely-Lellouch-Caprini parametrization \cite{prd79.013008}
  with the available experimental data and theoretical information
  and then extrapolating to the point of $q^{2}$ $=$ $0$,
  for example,
  $0.254^{+0.023}_{-0.022}$ in Ref. \cite{prd79.013008}\footnotemark[1],
  $0.248{\pm}0.082$ in Ref. \cite{prd91.074510},
  $0.20{\pm}0.14$ in Ref. \cite{prd92.014024},
  $0.254{\pm}0.081$ in Ref. \cite{prd97.054004}.
  Our results of $F_{0}^{B{\to}{\pi}}(0)$ are basically consistent
  those of Refs. \cite{prd79.013008,prd91.074510,prd92.014024,
  prd97.054004} within uncertainties.
  In addition, from the definition of formfactor in
  Eq.(\ref{formfactor-defination-01}),
  it is clear that there should be a relation between
  formfactors and decay constants,
  \begin{equation}
  \frac{F_{0}^{B{\to}{\pi}}(0)}{F_{0}^{B{\to}K}(0)}
  \, {\approx}\, \frac{f_{\pi}}{f_{K}}
  \label{hadronic-matrix-element}.
  \end{equation}
  The numbers in Table \ref{tab-formfactor} hold this relation
  well.
  The small violation arises from the $SU(3)$ flavor
  breaking effects.
  \footnotetext[1]{$F_{0}^{B{\to}{\pi}}(0)$ $=$ $f_{+}(0)$ is assumed.}

  \begin{figure}[ht]
  \includegraphics[width=0.2\textwidth,bb=180 535 320 650]{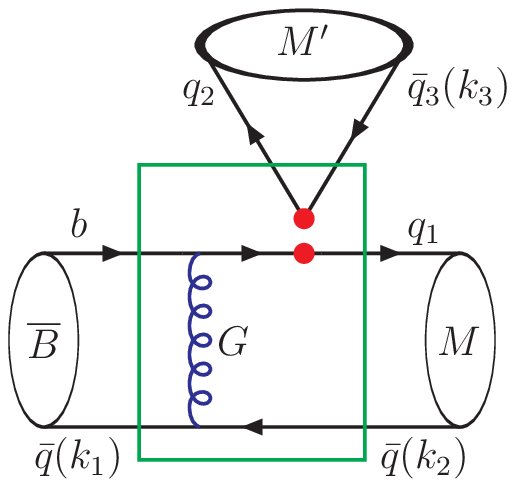}
  \quad
  \includegraphics[width=0.2\textwidth,bb=180 535 320 650]{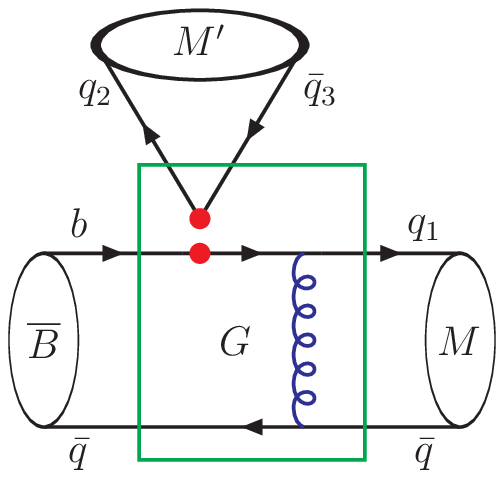}
  \quad
  \includegraphics[width=0.2\textwidth,bb=180 535 320 650]{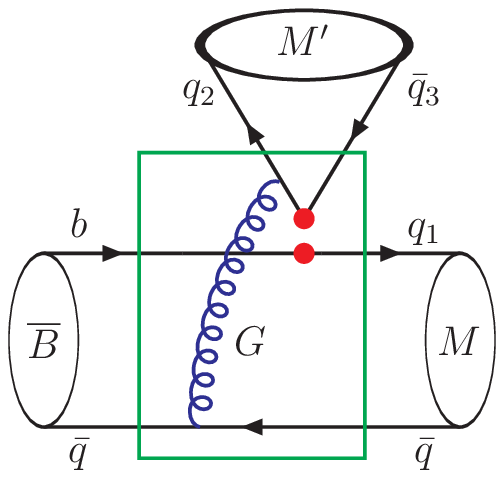}
  \quad
  \includegraphics[width=0.2\textwidth,bb=180 535 320 650]{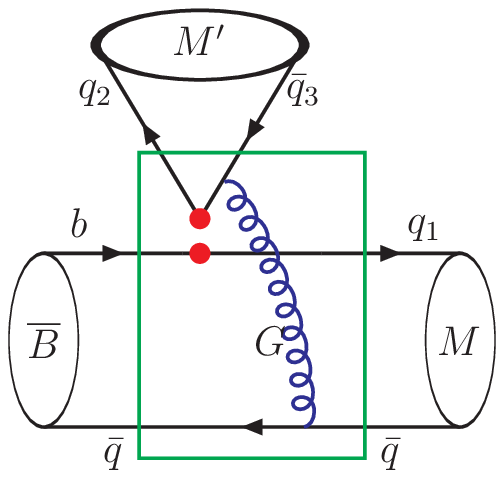}
  \\ {(a) \hspace{0.19\textwidth} (b) \hspace{0.19\textwidth}
      (c) \hspace{0.19\textwidth} (d) } \\
  \includegraphics[width=0.2\textwidth,bb=200 545 340 660]{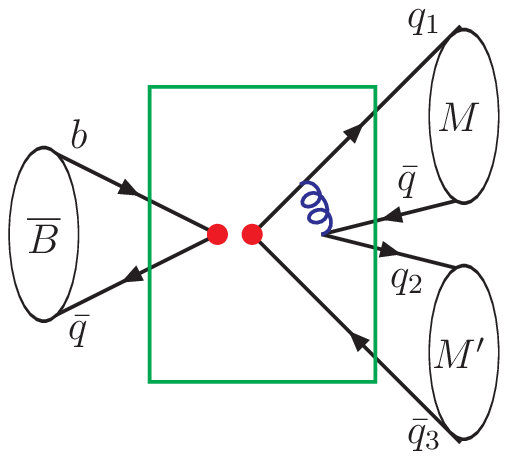}
  \quad
  \includegraphics[width=0.2\textwidth,bb=200 545 340 660]{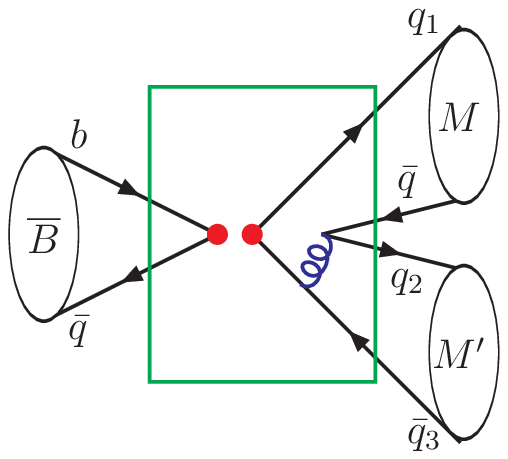}
  \quad
  \includegraphics[width=0.2\textwidth,bb=200 545 340 660]{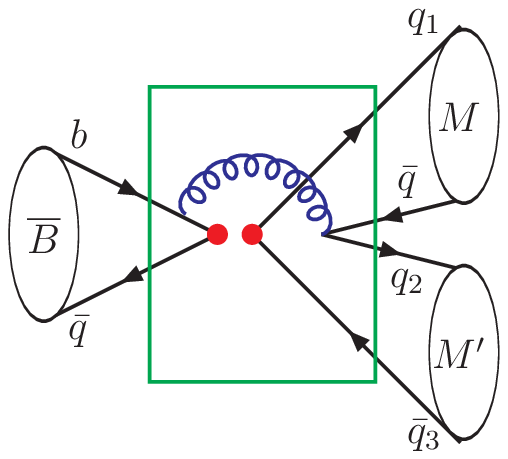}
  \quad
  \includegraphics[width=0.2\textwidth,bb=200 545 340 660]{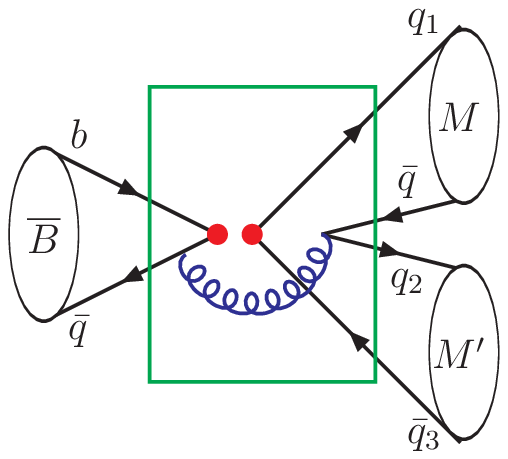}
  \\ {(e) \hspace{0.19\textwidth} (f) \hspace{0.19\textwidth}
      (g) \hspace{0.19\textwidth} (h) }
  \caption{Feynman diagrams contributing to the $\overline{B}$
   ${\to}$ $M\,M^{\prime}$ decays, where the dots denote appropriate
   interactions and boxes denote quark scattering amplitudes.
   (a) and (b) are factorizable emission diagrams.
   (c) and (d) are nonfactorizable emission diagrams.
   (e) and (f) are factorizable annihilation diagrams.
   (g) and (h) are nonfactorizable annihilation diagrams. }
  \label{fig-amplitude-01}
  \end{figure}

  The Feynman diagrams for two-body nonleptonic $B$ meson decays
  are shown in Fig. \ref{fig-amplitude-01}.
  The amplitudes $\mathcal{A}$ with the pQCD approach are usually
  divided into three parts : the short-distance contributions encoded
  in the Wilson coefficients $C_{i}$, the quark scattering amplitudes
  $\mathcal{\cal T}_{i}$, and hadron WFs ${\Phi}_{i}$.
  The general form of decay amplitude is
  \begin{equation}
  \mathcal{A}_{i}\ {\propto}\ {\int}\, {\prod_j}dx_{j}\,db_{j}\,
  C_{i}(t_{i})\, \mathcal{T}_{i}(t_{i},x_{j},b_{j})\,
  {\Phi}_{j}(x_{j},b_{j})\,e^{-S_{j}}
  \label{hadronic-matrix-element}.
  \end{equation}

  In the rest frame of the $B$ meson, the $CP$-averaged branching
  ratios are defined as :
   \begin{equation}
   \mathcal{B}r\, =\,
   \frac{{\tau}_{B}}{16{\pi}}\,
   \frac{p_{\rm cm}}{m_{B}^{2}}\, \big\{
  {\vert}{\cal A}(B{\to}f){\vert}^{2}+
  {\vert}{\cal A}(\overline{B}{\to}\bar{f}){\vert}^{2} \big\}
   \label{branching-ratio-definition},
   \end{equation}
  where ${\tau}_{B}$ is the lifetime of the $B$ meson.
  $p_{\rm cm}$ is the common momentum of final states.
  The decay amplitudes including the ${\phi}_{B2}$ contributions
  are listed in Appendix \ref{decay-modes}.
  For the charged $B_{u}$ meson decays, the $CP$ violating
  asymmetries arises from the interference between tree and
  penguin amplitudes. The direct $CP$ violating asymmetry
  is defined as follows.
   \begin{equation}
   \mathcal{A}_{CP}\, =\,
   \frac{{\Gamma}(B^{-}{\to}f)-{\Gamma}(B^{+}{\to}\bar{f})}
        {{\Gamma}(B^{-}{\to}f)+{\Gamma}(B^{+}{\to}\bar{f})}
   \, =\,
   \frac{{\vert}\mathcal{A}(B^{-}{\to}f){\vert}^{2}
        -{\vert}\mathcal{A}(B^{+}{\to}\bar{f}){\vert}^{2}}
        {{\vert}\mathcal{A}(B^{-}{\to}f){\vert}^{2}
        +{\vert}\mathcal{A}(B^{+}{\to}\bar{f}){\vert}^{2}}
   \label{direct-CP-definition-01}.
   \end{equation}
  For the neutral $B_{d}$ meson decays into final state $f$
  with $f$ $=$ $\bar{f}$, the time-dependent $CP$ violating
  asymmetry is defined as follows.
   \begin{equation}
   \mathcal{A}_{CP}\, =\,
   \frac{{\Gamma}(\bar{B}^{0}{\to}f)-{\Gamma}(B^{0}{\to}\bar{f})}
        {{\Gamma}(\bar{B}^{0}{\to}f)+{\Gamma}(B^{0}{\to}\bar{f})}
   \ {\approx}\
   S_{f}\,{\sin}(x\,{\Gamma}\,t)-C_{f}\,{\cos}(x\,{\Gamma}\,t)
   \label{CP-asymmetry-definition-02},
   \end{equation}
  with the $y$ $=$
  ${\displaystyle \frac{{\Delta}{\Gamma}}{2\,{\Gamma}_{B}} }$
  ${\simeq}$ $0$ approximation, where $x$ $=$
  ${\displaystyle \frac{{\Delta}m_{B}}{{\Gamma}_{B}} }$ $=$
  $0.769(4)$ \cite{pdg2020} is the $B^{0}$-$\overline{B}^{0}$
  oscillation parameter.
  ${\Gamma}_{B}$ $=$ ${\displaystyle \frac{1}{{\tau}_{B}} }$
  is the full width of the $B_{d}$ meson.
  $C_{f}$ and $S_{f}$ are the direct and mixing-induced
  $CP$ asymmetries.
   \begin{equation}
   C_{f}\, =\, \frac{ 1-{\vert}{\lambda}_{f}{\vert}^{2} }
                    { 1+{\vert}{\lambda}_{f}{\vert}^{2} }
   \label{CP-asymmetry-definition-03},
   \end{equation}
   \begin{equation}
   S_{f}\, =\, \frac{ 2\,{\cal I}m({\lambda}_{f}) }
                    { 1+{\vert}{\lambda}_{f}{\vert}^{2} }
   \label{CP-asymmetry-definition-04},
   \end{equation}
   \begin{equation}
  {\lambda}_{f}\, =\,
   \frac{ V_{tb}^{\ast}\,V_{td} }{ V_{tb}\,V_{td}^{\ast} }\,
   \frac{ \mathcal{A}(\bar{B}^{0}{\to}f) }
        { \mathcal{A}(B^{0}{\to}f) }
   \label{CP-asymmetry-definition-05}.
   \end{equation}

  \begin{table}[ht]
  \caption{The $CP$-averaged branching ratios ${\cal B}r$ (in
  the unit of $10^{-6}$) for the $B_{u}$ ${\to}$ $PP$ decays.
  The theoretical results are respectively calculated with
  parameters of S1, S2, S3 scenarios to compare with data of
  PDG, Belle and BaBar \cite{pdg2020}.
  The theoretical uncertainties arise from
  variations of ${\omega}_{B}{\pm}0.01$ GeV,
  ${\mu}_{M}{\pm}0.1$ GeV, and $a_{2}^{M}{\pm}0.15$, respectively.}
  \label{table-branching-ratio-bu}
  \begin{ruledtabular}
  \begin{tabular}{cccccc}
  & mode
  & $B^{-}$ ${\to}$ ${\pi}^{-}{\pi}^{0}$
  & $B^{-}$ ${\to}$ ${\pi}^{-}\overline{K}^{0}$
  & $B^{-}$ ${\to}$ ${\pi}^{0}K^{-}$
  & $B^{-}$ ${\to}$ $K^{0}K^{-}$ \\ \hline
  & PDG
  & $5.5{\pm}0.4$
  & $23.7{\pm}0.8$
  & $12.9{\pm}0.5$
  & $1.31{\pm}0.17$ \\
    S1 & ${\phi}_{B1}$+${\phi}_{B2}$
  & $  3.04^{+  0.18+  0.16+  0.27}_{  -0.17  -0.16  -0.25}$
  & $ 24.12^{+  1.73+  3.69+  2.50}_{  -1.60  -3.32  -2.36}$
  & $ 13.13^{+  0.93+  1.95+  1.11}_{  -0.87  -1.76  -1.07}$
  & $  1.95^{+  0.13+  0.34+  0.18}_{  -0.13  -0.30  -0.16}$ \\
  & ${\phi}_{B1}$
  & $  2.16^{+  0.14+  0.15+  0.14}_{  -0.13  -0.15  -0.13}$
  & $ 17.89^{+  1.36+  2.97+  1.30}_{  -1.26  -2.65  -1.24}$
  & $  9.71^{+  0.73+  1.57+  0.58}_{  -0.68  -1.41  -0.56}$
  & $  1.45^{+  0.11+  0.27+  0.07}_{  -0.10  -0.25  -0.06}$ \\ \hline
  & Belle
  & $5.86{\pm}0.46$
  & $23.97{\pm}0.89$
  & $12.62{\pm}0.64$
  & $1.11{\pm}0.20$ \\
    S2 & ${\phi}_{B1}$+${\phi}_{B2}$
  & $  3.23^{+  0.19+  0.18+  0.29}_{  -0.18  -0.18  -0.27}$
  & $ 23.87^{+  1.72+  3.83+  2.51}_{  -1.60  -3.43  -2.38}$
  & $ 13.03^{+  0.93+  2.03+  1.12}_{  -0.87  -1.83  -1.07}$
  & $  1.88^{+  0.13+  0.35+  0.17}_{  -0.12  -0.31  -0.15}$ \\
  &  ${\phi}_{B1}$
  & $  2.27^{+  0.15+  0.17+  0.15}_{  -0.14  -0.16  -0.14}$
  & $ 17.64^{+  1.36+  3.08+  1.30}_{  -1.25  -2.74  -1.25}$
  & $  9.60^{+  0.74+  1.64+  0.59}_{  -0.68  -1.46  -0.57}$
  & $  1.38^{+  0.10+  0.29+  0.06}_{  -0.10  -0.25  -0.05}$ \\ \hline
  & BaBar
  & $5.02{\pm}0.54$
  & $23.9{\pm}1.5$
  & $13.6{\pm}0.9$
  & $1.61{\pm}0.45$ \\
    S3 & ${\phi}_{B1}$+${\phi}_{B2}$
  & $  3.42^{+  0.21+  0.20+  0.32}_{  -0.19  -0.20  -0.30}$
  & $ 23.49^{+  1.71+  3.96+  2.51}_{  -1.58  -3.53  -2.38}$
  & $ 12.87^{+  0.93+  2.11+  1.12}_{  -0.86  -1.89  -1.07}$
  & $  1.79^{+  0.12+  0.36+  0.15}_{  -0.12  -0.32  -0.13}$ \\
  & ${\phi}_{B1}$
  & $  2.39^{+  0.16+  0.18+  0.16}_{  -0.14  -0.18  -0.15}$
  & $ 17.30^{+  1.35+  3.19+  1.30}_{  -1.24  -2.81  -1.24}$
  & $  9.44^{+  0.73+  1.70+  0.59}_{  -0.67  -1.51  -0.57}$
  & $  1.30^{+  0.10+  0.29+  0.05}_{  -0.09  -0.26  -0.04}$
  \end{tabular}
  \end{ruledtabular}
  \end{table}
  \begin{table}[ht]
  \caption{The numerical values of the $CP$-averaged branching
  ratios (in the unit of $10^{-6}$) for the $B_{d}$ ${\to}$ $PP$
  decays. Other legends are the same as those of Table.
  \ref{table-branching-ratio-bu}.}
  \label{table-branching-ratio-bd}
  \begin{ruledtabular}
  \begin{tabular}{ccccc}
  &  mode
  & $\overline{B}^{0}$ ${\to}$ ${\pi}^{+}{\pi}^{-}$
  & $\overline{B}^{0}$ ${\to}$ ${\pi}^{0}{\pi}^{0}$
  & $\overline{B}^{0}$ ${\to}$ ${\pi}^{+}K^{-}$ \\ \hline
  & PDG
  & $5.12{\pm}0.19$
  & $1.59{\pm}0.26$
  & $19.6{\pm}0.5$ \\
    S1 & ${\phi}_{B1}$+${\phi}_{B2}$
  & $  5.52^{+  0.34+  0.37+  0.47}_{  -0.32  -0.36  -0.43}$
  & $  0.24^{+  0.02+  0.04+  0.01}_{  -0.02  -0.03  -0.01}$
  & $ 20.07^{+  1.44+  3.13+  2.05}_{  -1.33  -2.81  -1.94}$ \\
  & ${\phi}_{B1}$
  & $  3.81^{+  0.25+  0.31+  0.23}_{  -0.23  -0.30  -0.20}$
  & $  0.19^{+  0.01+  0.03+  0.02}_{  -0.01  -0.03  -0.01}$
  & $ 15.00^{+  1.14+  2.53+  1.07}_{  -1.05  -2.25  -1.02}$ \\ \hline
  & Belle
  & $5.04{\pm}0.28$
  & $1.31{\pm}0.27$
  & $20.00{\pm}0.69$ \\
    S2 & ${\phi}_{B1}$+${\phi}_{B2}$
  & $  5.82^{+  0.36+  0.41+  0.50}_{  -0.34  -0.39  -0.46}$
  & $  0.24^{+  0.02+  0.04+  0.01}_{  -0.02  -0.04  -0.01}$
  & $ 19.81^{+  1.43+  3.24+  2.05}_{  -1.33  -2.89  -1.95}$ \\
  & ${\phi}_{B1}$
  & $  3.99^{+  0.26+  0.34+  0.24}_{  -0.25  -0.33  -0.21}$
  & $  0.18^{+  0.01+  0.03+  0.02}_{  -0.01  -0.03  -0.01}$
  & $ 14.76^{+  1.14+  2.62+  1.07}_{  -1.05  -2.32  -1.02}$ \\ \hline
  & BaBar
  & $5.5{\pm}0.5$
  & $1.83{\pm}0.25$
  & $19.1{\pm}0.8$ \\
    S3 & ${\phi}_{B1}$+${\phi}_{B2}$
  & $  6.12^{+  0.39+  0.45+  0.54}_{  -0.36  -0.43  -0.50}$
  & $  0.23^{+  0.02+  0.04+  0.01}_{  -0.02  -0.04  -0.01}$
  & $ 19.45^{+  1.42+  3.34+  2.04}_{  -1.31  -2.97  -1.94}$ \\
  & ${\phi}_{B1}$
  & $  4.16^{+  0.28+  0.37+  0.26}_{  -0.26  -0.36  -0.23}$
  & $  0.18^{+  0.01+  0.03+  0.02}_{  -0.01  -0.03  -0.01}$
  & $ 14.43^{+  1.13+  2.70+  1.06}_{  -1.04  -2.38  -1.01}$ \\ \hline \hline
  & mode
  & $\overline{B}^{0}$ ${\to}$ ${\pi}^{0}\overline{K}^{0}$
  & $\overline{B}^{0}$ ${\to}$ $K^{+}K^{-}$
  & $\overline{B}^{0}$ ${\to}$ $K^{0}\overline{K}^{0}$ \\ \hline
  & PDG
  & $9.9{\pm}0.5$
  & $0.078{\pm}0.015$
  & $1.21{\pm}0.16$ \\
    S1 & ${\phi}_{B1}$+${\phi}_{B2}$
  & $  9.11^{+  0.66+  1.48+  1.15}_{  -0.61  -1.32  -1.08}$
  & $ 0.107^{+ 0.003+ 0.005+ 0.062}_{ -0.003 -0.004 -0.048}$
  & $  1.43^{+  0.10+  0.25+  0.05}_{  -0.09  -0.23  -0.03}$ \\
  & ${\phi}_{B1}$
  & $  6.81^{+  0.52+  1.19+  0.61}_{  -0.48  -1.05  -0.57}$
  & $ 0.089^{+ 0.003+ 0.004+ 0.047}_{ -0.003 -0.004 -0.037}$
  & $  1.06^{+  0.08+  0.21+  0.03}_{  -0.07  -0.19  -0.01}$ \\ \hline
  & Belle
  & $9.68{\pm}0.68$
  & $0.10{\pm}0.09$
  & $1.26{\pm}0.20$  \\
    S2 & ${\phi}_{B1}$+${\phi}_{B2}$
  & $  8.95^{+  0.65+  1.53+  1.15}_{  -0.61  -1.36  -1.08}$
  & $ 0.109^{+ 0.003+ 0.005+ 0.065}_{ -0.003 -0.004 -0.050}$
  & $  1.37^{+  0.10+  0.26+  0.03}_{  -0.09  -0.24  -0.02}$ \\
  & ${\phi}_{B1}$
  & $  6.67^{+  0.52+  1.22+  0.60}_{  -0.48  -1.08  -0.57}$
  & $ 0.091^{+ 0.003+ 0.004+ 0.050}_{ -0.003 -0.004 -0.039}$
  & $  1.01^{+  0.07+  0.22+  0.04}_{  -0.07  -0.19  -0.02}$ \\ \hline
  & BaBar
  & $10.1{\pm}0.7$
  &
  & $1.08{\pm}0.30$ \\
    S3 & ${\phi}_{B1}$+${\phi}_{B2}$
  & $  8.75^{+  0.64+  1.57+  1.15}_{  -0.60  -1.38  -1.07}$
  & $ 0.111^{+ 0.003+ 0.005+ 0.068}_{ -0.003 -0.004 -0.052}$
  & $  1.30^{+  0.09+  0.27+  0.02}_{  -0.08  -0.24  -0.01}$ \\
  & ${\phi}_{B1}$
  & $  6.50^{+  0.51+  1.26+  0.60}_{  -0.47  -1.10  -0.57}$
  & $ 0.093^{+ 0.003+ 0.004+ 0.053}_{ -0.003 -0.004 -0.040}$
  & $  0.94^{+  0.07+  0.22+  0.05}_{  -0.06  -0.20  -0.03}$
  \end{tabular}
  \end{ruledtabular}
  \end{table}
  \begin{table}[ht]
  \caption{The $CP$ asymmetries $\mathcal{A}_{CP}$ (in the unit
  of $10^{-2}$) for $B$ ${\to}$ $PP$ decays. Other legends are
  the same as those of Table \ref{table-branching-ratio-bu}.}
  \label{table-cp-bu}
  \begin{ruledtabular} {\footnotesize
  \begin{tabular}{cccccc}
    $\mathcal{A}_{CP}$
  & ${\pi}^{-}{\pi}^{0}$
  & ${\pi}^{-}\overline{K}^{0}$
  & ${\pi}^{0}K^{-}$
  & $K^{0}K^{-}$
  & ${\pi}^{+}K^{-}$ \\ \hline
    PDG
  & $3{\pm}4$
  & $-1.7{\pm}1.6$
  & $3.7{\pm}2.1$
  & $4{\pm}14$
  & $-8.3{\pm}0.4$ \\
    ${\phi}_{B1}$+${\phi}_{B2}$
  & $-0.004$
  & $ -0.67^{+  0.02+  0.04+  0.14}_{  -0.02  -0.05  -0.11}$
  & $ -6.25^{+  0.25+  0.30+  1.51}_{  -0.26  -0.33  -1.56}$
  & $ 14.78^{+  0.27+  1.05+  3.33}_{  -0.27  -0.90  -3.87}$
  & $ -7.33^{+  0.29+  0.41+  1.98}_{  -0.30  -0.44  -2.04}$ \\
    ${\phi}_{B1}$
  & $-0.02$
  & $ -0.74^{+  0.02+  0.04+  0.14}_{  -0.02  -0.05  -0.10}$
  & $ -6.73^{+  0.30+  0.36+  1.74}_{  -0.31  -0.40  -1.88}$
  & $ 14.83^{+  0.26+  1.10+  3.28}_{  -0.26  -0.93  -3.71}$
  & $ -8.14^{+  0.33+  0.49+  2.38}_{  -0.34  -0.54  -2.54}$ \\ \hline
    Belle
  & $2.5{\pm}4.4$
  & $-1.1{\pm}2.2$
  & $4.3{\pm}2.4$
  & $1.4{\pm}16.8$
  & $-6.9{\pm}1.6$ \\
    ${\phi}_{B1}$+${\phi}_{B2}$
  & $-0.01$
  & $ -0.69^{+  0.02+  0.05+  0.15}_{  -0.02  -0.05  -0.12}$
  & $ -6.05^{+  0.25+  0.30+  1.43}_{  -0.26  -0.33  -1.49}$
  & $ 15.28^{+  0.28+  1.21+  3.53}_{  -0.28  -1.03  -4.08}$
  & $ -7.18^{+  0.28+  0.41+  1.91}_{  -0.29  -0.45  -1.97}$ \\
    ${\phi}_{B1}$
  & $-0.02$
  & $ -0.75^{+  0.02+  0.05+  0.15}_{  -0.02  -0.05  -0.12}$
  & $ -6.51^{+  0.29+  0.36+  1.66}_{  -0.30  -0.40  -1.80}$
  & $ 15.39^{+  0.27+  1.30+  3.55}_{  -0.27  -1.08  -3.98}$
  & $ -7.98^{+  0.33+  0.50+  2.31}_{  -0.34  -0.54  -2.47}$ \\ \hline
    BaBar
  & $3{\pm}8$
  & $-2.9{\pm}4.0$
  & $3{\pm}4$
  & $10{\pm}26$
  & $-10.7{\pm}1.7$ \\
    ${\phi}_{B1}$+${\phi}_{B2}$
  & $-0.01$
  & $ -0.70^{+  0.02+  0.05+  0.16}_{  -0.02  -0.06  -0.13}$
  & $ -5.86^{+  0.24+  0.30+  1.36}_{  -0.25  -0.32  -1.41}$
  & $ 15.90^{+  0.29+  1.41+  3.78}_{  -0.29  -1.18  -4.34}$
  & $ -7.03^{+  0.28+  0.41+  1.84}_{  -0.29  -0.44  -1.89}$ \\
    ${\phi}_{B1}$
  & $-0.03$
  & $ -0.77^{+  0.02+  0.05+  0.16}_{  -0.02  -0.06  -0.13}$
  & $ -6.30^{+  0.29+  0.36+  1.57}_{  -0.30  -0.40  -1.72}$
  & $ 16.12^{+  0.29+  1.56+  3.89}_{  -0.29  -1.27  -4.31}$
  & $ -7.83^{+  0.33+  0.50+  2.24}_{  -0.34  -0.55  -2.40}$
  \end{tabular} }
  \end{ruledtabular}
  \end{table}
  \begin{table}[ht]
  \caption{The $CP$ asymmetries (in the unit of $10^{-2}$) for
  $B_{d}$ ${\to}$ $PP$ decays. Other legends are the same as
  those of Table \ref{table-branching-ratio-bu}.}
  \label{table-cp-bd}
  \begin{ruledtabular}
  \begin{tabular}{ccccc}
  & $C_{{\pi}^{+}{\pi}^{-}}$
  & $S_{{\pi}^{+}{\pi}^{-}}$
  & $C_{{\pi}^{0}{\pi}^{0}}$
  & $S_{{\pi}^{0}{\pi}^{0}}$ \\ \hline
    PDG
  & $-32{\pm}4$
  & $-65{\pm}4$
  & $-33{\pm}22$
  & \\
    ${\phi}_{B1}$+${\phi}_{B2}$
  & $-17.85^{+  0.55+  0.05+  3.84}_{  -0.54  -0.02  -3.42}$
  & $-81.91^{+  0.17+  1.57+  2.43}_{  -0.17  -1.52  -1.96}$
  & $ 38.34^{+  1.39+  1.78+  5.37}_{  -1.37  -1.57  -5.40}$
  & $ 89.03^{+  0.49+  1.13+  0.97}_{  -0.51  -1.38  -1.21}$ \\
    ${\phi}_{B1}$
  & $-21.52^{+  0.71+  0.13+  4.96}_{  -0.69  -0.10  -4.52}$
  & $-80.15^{+  0.15+  1.65+  2.94}_{  -0.16  -1.60  -2.30}$
  & $ 39.06^{+  1.60+  1.86+  4.49}_{  -1.58  -1.62  -4.17}$
  & $ 85.90^{+  0.53+  1.32+  0.35}_{  -0.54  -1.59  -0.25}$ \\ \hline
    Belle
  & $-33{\pm}7$
  & $-64{\pm}9$
  & $-14{\pm}37$
  &  \\
    ${\phi}_{B1}$+${\phi}_{B2}$
  & $-16.75^{+  0.51+  0.05+  3.52}_{  -0.52  -0.02  -3.11}$
  & $-83.12^{+  0.16+  1.55+  2.34}_{  -0.16  -1.50  -1.91}$
  & $ 37.29^{+  1.41+  1.91+  5.29}_{  -1.40  -1.68  -5.09}$
  & $ 88.69^{+  0.50+  1.31+  0.83}_{  -0.51  -1.63  -1.09}$ \\
    ${\phi}_{B1}$
  & $-20.24^{+  0.66+  0.12+  4.57}_{  -0.68  -0.09  -4.13}$
  & $-81.46^{+  0.15+  1.64+  2.86}_{  -0.15  -1.59  -2.28}$
  & $ 37.65^{+  1.63+  2.01+  4.49}_{  -1.62  -1.75  -3.86}$
  & $ 85.41^{+  0.54+  1.54+  0.34}_{  -0.55  -1.88  -0.28}$ \\ \hline
    BaBar
  & $-25{\pm}8$
  & $-68{\pm}10$
  & $-43{\pm}26$
  & \\
    ${\phi}_{B1}$+${\phi}_{B2}$
  & $-15.72^{+  0.48+  0.05+  3.23}_{  -0.49  -0.02  -2.82}$
  & $-84.33^{+  0.15+  1.53+  2.24}_{  -0.15  -1.48  -1.85}$
  & $ 36.32^{+  1.43+  2.08+  5.21}_{  -1.42  -1.80  -4.74}$
  & $ 88.11^{+  0.50+  1.55+  0.70}_{  -0.51  -1.96  -1.00}$ \\
    ${\phi}_{B1}$
  & $-19.04^{+  0.63+  0.12+  4.20}_{  -0.64  -0.09  -3.76}$
  & $-82.79^{+  0.14+  1.62+  2.77}_{  -0.14  -1.58  -2.24}$
  & $ 36.33^{+  1.66+  2.22+  4.47}_{  -1.63  -1.89  -3.51}$
  & $ 84.65^{+  0.55+  1.83+  0.37}_{  -0.56  -2.27  -0.37}$ \\ \hline \hline
  & $C_{{\pi}^{0}\overline{K}^{0}}$
  & $S_{{\pi}^{0}\overline{K}^{0}}$
  & $C_{K^{+}K^{-}}$
  & $S_{K^{+}K^{-}}$ \\ \hline
    PDG
  & $0{\pm}13$
  & $58{\pm}17$ \\
    ${\phi}_{B1}$+${\phi}_{B2}$
  & $ -0.54^{+  0.01+  0.05+  0.25}_{  -0.01  -0.05  -0.21}$
  & $ 68.74^{+  0.02+  0.16+  0.20}_{  -0.02  -0.13  -0.16}$
  & $-81.17^{+  0.45+  0.79+  0.13}_{  -0.45  -0.70  -2.04}$
  & $-28.21^{+  0.94+  1.59+  0.95}_{  -0.93  -1.58  -0.53}$ \\
    ${\phi}_{B1}$
  & $ -0.85^{+  0.01+  0.06+  0.36}_{  -0.01  -0.06  -0.31}$
  & $ 69.29^{+  0.03+  0.18+  0.13}_{  -0.03  -0.15  -0.14}$
  & $-82.90^{+  0.46+  0.67+  0.27}_{  -0.45  -0.59  -1.24}$
  & $-28.35^{+  1.02+  1.25+  0.35}_{  -1.01  -1.27  -1.24}$ \\ \hline
    Belle
  & $-14{\pm}14$
  & $67{\pm}32$  \\
    ${\phi}_{B1}$+${\phi}_{B2}$
  & $ -0.60^{+  0.01+  0.05+  0.27}_{  -0.00  -0.06  -0.23}$
  & $ 68.86^{+  0.02+  0.18+  0.20}_{  -0.02  -0.15  -0.16}$
  & $-81.26^{+  0.44+  0.88+  0.11}_{  -0.43  -0.78  -1.96}$
  & $-27.97^{+  0.94+  1.66+  0.24}_{  -0.92  -1.65  -0.12}$ \\
    ${\phi}_{B1}$
  & $ -0.93^{+  0.01+  0.07+  0.39}_{  -0.01  -0.07  -0.34}$
  & $ 69.40^{+  0.04+  0.21+  0.15}_{  -0.04  -0.18  -0.16}$
  & $-83.11^{+  0.44+  0.76+  0.25}_{  -0.43  -0.67  -1.24}$
  & $-27.65^{+  1.01+  1.35+  0.75}_{  -0.99  -1.37  -1.80}$ \\ \hline
    BaBar
  & $13{\pm}13$
  & $55{\pm}20$ \\
    ${\phi}_{B1}$+${\phi}_{B2}$
  & $ -0.67^{+  0.00+  0.06+  0.30}_{  -0.00  -0.07  -0.25}$
  & $ 68.99^{+  0.02+  0.21+  0.20}_{  -0.02  -0.18  -0.16}$
  & $-81.24^{+  0.42+  0.97+  0.08}_{  -0.41  -0.87  -1.81}$
  & $-27.80^{+  0.93+  1.71+  0.35}_{  -0.92  -1.70  -0.56}$ \\
    ${\phi}_{B1}$
  & $ -1.02^{+  0.01+  0.08+  0.43}_{  -0.00  -0.09  -0.36}$
  & $ 69.53^{+  0.04+  0.24+  0.16}_{  -0.04  -0.20  -0.17}$
  & $-83.19^{+  0.41+  0.86+  0.25}_{  -0.40  -0.76  -1.24}$
  & $-27.06^{+  1.00+  1.45+  1.18}_{  -0.99  -1.46  -2.44}$
  \end{tabular}
  \end{ruledtabular}
  \end{table}
  \begin{figure}[ht]
  \includegraphics[width=0.5\textwidth]{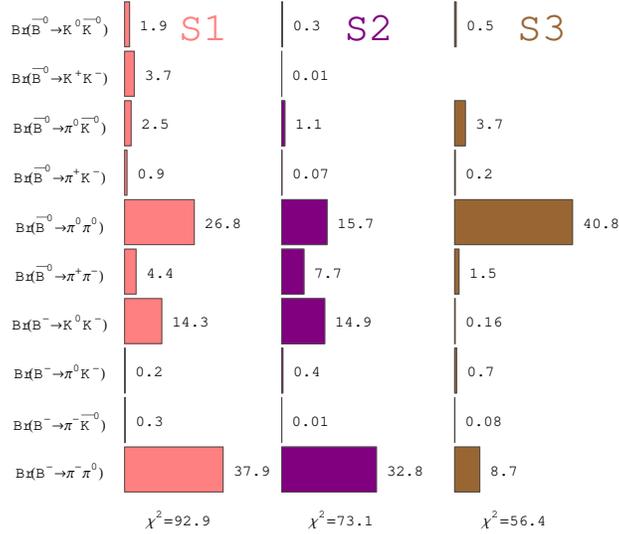}
  \caption{The ${\chi}^{2}$ distribution of branching ratios
  for three optimal scenarios.}
  \label{br-chi}
  \end{figure}

  The numerical results on the $CP$-averaged branching ratios
  together with experimental data are presented in Table
  \ref{table-branching-ratio-bu} and \ref{table-branching-ratio-bd},
  $CP$ asymmetries in Table \ref{table-cp-bu} and \ref{table-cp-bd}.
  Using the minimum ${\chi}^{2}$ method,
   \begin{equation}
  {\chi}^{2}\, =\, \sum\limits_{i}
   \frac{ ({\cal B}r^{\rm theo.}-{\cal B}r^{\rm exp.})^{2} }
        { {\sigma}_{{\cal B}r^{\rm exp.}}^{2} }
   \label{chi2},
   \end{equation}
  three optimal scenarios (S1, S2 and S3) of parameters
  ${\omega}_{B}$ and ${\mu}_{M}$ are obtained when the
  contributions of ${\phi}_{B2}$ are considered.
  For the ten concerned $B$ decay modes, ${\chi}^{2}/d.o.f$
  $=$ $92.9/8$, $73.1/8$ and $56.4/7$ correspond to data of
  PDG, Belle and BaBar, respectively.
  The agreement between theoretical and experimental
  results is illustrated by the ${\chi}^{2}$ distribution
  in Fig. \ref{br-chi}.
  The followings are our comments.

  (1) From Table \ref{table-branching-ratio-bu} and
  \ref{table-branching-ratio-bd}, it is seen that the contributions
  of WF ${\phi}_{B2}$ are more than 25\% of total branching ratios,
  except for the pure annihilation $\overline{B}^{0}$ ${\to}$
  $K^{+}K^{-}$ decay.
  That is because WF ${\phi}_{B2}$ contributes nothing to the
  factorizable annihilation amplitudes of
  Eqs.(\ref{amp-a01-01})-(\ref{amp-a02-03}).
  From Table \ref{table-cp-bu} and \ref{table-cp-bd}, it is seen
  that the contributions of WF ${\phi}_{B2}$ result in a small
  reduction of direct $CP$ asymmetries.

  (2) From appendix \ref{decay-modes}, it is clearly seen that for
  the $B$ ${\to}$ ${\pi}{\pi}$ and $\overline{K}K$ decays, the CKM
  factors of the tree and penguin amplitudes are respectively
  $V_{ub}\,V_{ud}^{\ast}$ and $V_{tb}\,V_{td}^{\ast}$, and
  have the same order of magnitude ${\propto}$ ${\lambda}^{3}$.
  For the $B$ ${\to}$ ${\pi}K$ decays, the tree amplitudes
  being proportional to the CKM factor $V_{ub}\,V_{us}^{\ast}$
  are suppressed by ${\lambda}^{2}$ compared with the penguin
  amplitudes being proportional to the CKM factor
  $V_{tb}\,V_{ts}^{\ast}$. In addition,
  the theoretical and experimental results in Table
  \ref{table-branching-ratio-bu} and \ref{table-branching-ratio-bd}
  show that branching ratios for $B$ ${\to}$ ${\pi}K$ decays are
  in general larger than those for $B$ ${\to}$ ${\pi}{\pi}$
  and $\overline{K}K$ decays.
  These facts confirm previous studies \cite{npb675.333,plb504.6}
  that penguin contributions are dynamically enhanced and essential
  for explaining the $B$ ${\to}$ ${\pi}K$ decays.
  What is more, our studies show that the nonfactorizable
  annihilation amplitudes mainly from WF ${\phi}_{B1}$ rather than
  ${\phi}_{B2}$ provide large strong phases for the $B$ ${\to}$
  ${\pi}K$ decays, as analyzed in Ref. \cite{plb504.6}.

  (3) From Fig. \ref{br-chi}, it is seen that (i) for the $B$ ${\to}$
  ${\pi}K$ decays, when the contributions of WF ${\phi}_{B2}$ are
  included, theoretical results of branching ratio can give a
  satisfactory explanation on experimental data.
  Compared the numbers in Table \ref{table-branching-ratio-bu} and
  \ref{table-branching-ratio-bd} with the NLO results of Refs.
  \cite{prd72.114005,prd90.074018,prd93.014024}
  (see Table \ref{previous-b-pi-k}), it is seen that
  the contributions of ${\phi}_{B2}$ to branching ratios at the
  leading order (LO) is roughly equivalent to the NLO corrections
  without the participation of ${\phi}_{B2}$.
  (ii) The consideration of WF ${\phi}_{B2}$ cannot well settle the
  so-called $CP$ asymmetries ``$K{\pi}$'' puzzle, {\em i.e.}
  the discrepancy between theoretical and experimental results
  of ${\cal A}_{CP}(B^{-}{\to}{\pi}^{0}K^{-})$ $-$
  ${\cal A}_{CP}(\overline{B}^{0}{\to}{\pi}^{+}K^{-})$.
  The studies of Refs. \cite{prd90.074018,prd93.014024} showed
  that the NLO corrections including the glauber effects could
  flip the sign of ${\cal A}_{CP}(B^{-}{\to}{\pi}^{0}K^{-})$.
  It should be noted that the NLO and NLOG theoretical
  uncertainties of branching ratios are still large, and the
  current measurement accuracy of
  ${\cal A}_{CP}(B^{-}{\to}{\pi}^{0}K^{-})$ needs to be
  improved.
  \begin{table}[ht]
  \caption{The previous LO and NLO pQCD results for the $B$ ${\to}$
  ${\pi}K$ decays, where NLO and NLOG denote without and with the
  glauber effects, the unit of branching ratios and direct $CP$
  asymmetries are respectively $10^{-6}$ and $10^{-2}$.}
  \label{previous-b-pi-k}
  \begin{ruledtabular}
  \begin{tabular}{cccccc}
    mode & LO \cite{prd72.114005}
         & NLO \cite{prd72.114005}
         & NLOG\footnotemark[2] \cite{prd90.074018}
         & NLO \cite{prd93.014024}
         & NLOG \cite{prd93.014024} \\ \hline
    ${\cal B}r(B^{-}{\to}{\pi}^{-}\overline{K}^{0})$
  & $17.0$ & $24.5^{+13.6}_{-~8.1}$ & $21.1$
  & $27.2^{+9.3}_{-6.7}$ & $24.1^{+8.3}_{-6.0}$\\
    ${\cal B}r(B^{-}{\to}{\pi}^{0}K^{-})$
  & $10.2$ & $13.9^{+10.0}_{-~5.6}$ & $12.9$
  & $15.3^{+5.2}_{-3.8}$ & $14.0^{+4.7}_{-3.5}$ \\
    ${\cal B}r(\overline{B}^{0}{\to}{\pi}^{+}K^{-})$
  & $14.2$ & $20.9^{+15.6}_{-~8.3}$ & $17.7$
  & $23.3^{+7.8}_{-5.7}$ & $21.7^{+7.4}_{-5.3}$ \\
    ${\cal B}r(\overline{B}^{0}{\to}{\pi}^{0}\overline{K}^{0})$
  & $5.7$ & $9.1^{+5.6}_{-3.3}$ & $7.2$
  & $10.2^{+3.4}_{-2.5}$ & $9.3^{+3.2}_{-2.3}$ \\ \hline
    ${\cal A}_{CP}(B^{-}{\to}{\pi}^{0}K^{-})$
  & $-8$ & $-1^{+3}_{-5}$ & $10$ & $-0.8^{+1.3}_{-1.4}$ & $2.1{\pm}1.6$ \\
    ${\cal A}_{CP}(\overline{B}^{0}{\to}{\pi}^{+}K^{-})$
  & $-12$ & $-9^{+6}_{-8}$ & $-11$ & $-7.6{\pm}1.7$ & $-8.1{\pm}1.7$
  \end{tabular}
  \end{ruledtabular}
  \footnotetext[2]{The glauber phases $S_{e}$ $=$ $S_{e1}$ $=$
  $S_{e2}$ $=$ ${\displaystyle -\frac{\pi}{2} }$ is assumed.}
  \end{table}

  (4) From Fig. \ref{br-chi} and Table \ref{table-branching-ratio-bu}
  and \ref{table-branching-ratio-bd}, it is seen that for the
  $B$ ${\to}$ ${\pi}{\pi}$ decays, the pQCD results of branching
  ratios deviate from the current experimental measurement.
  The contributions of WF ${\phi}_{B2}$ can enhance the branching
  ratios and reduce these deviations.
  Compared the numbers in Table \ref{table-branching-ratio-bu} and
  \ref{table-branching-ratio-bd} with the NLO results of Refs.
  \cite{prd72.114005,prd90.074018,fp.16.24201,prd90.014029,prd91.114019}
  (see Table \ref{previous-b-pi-pi}), it is seen that (i) for the
  $\overline{B}^{0}$ ${\to}$ ${\pi}{\pi}$ decays, the LO
  contributions of ${\phi}_{B2}$ to branching ratios is
  roughly equivalent to the NLO corrections without the
  participation of ${\phi}_{B2}$. (ii) Besides the large
  theoretical uncertainties, the pQCD results,
  including either WF ${\phi}_{B2}$ or the NLO contributions,
  cannot well explain data on branching ratio for the
  $\overline{B}^{0}$ ${\to}$ ${\pi}^{0}{\pi}^{0}$ decay
  and $CP$ asymmetries for the $\overline{B}^{0}$ ${\to}$
  ${\pi}^{+}{\pi}^{-}$ decay.
  \begin{table}[ht]
  \caption{The previous LO and NLO pQCD results for the $B$ ${\to}$
  ${\pi}{\pi}$ decays. Other legends are the same as those of
  Table \ref{previous-b-pi-k}.}
  \label{previous-b-pi-pi}
  \begin{ruledtabular}
  \begin{tabular}{cccccccc}
    mode & LO \cite{prd72.114005}
         & NLO \cite{prd72.114005}
         & NLOG\footnotemark[3] \cite{prd90.074018}
         & LO \cite{prd90.014029}
         & NLO \cite{prd90.014029}
         & NLO \cite{prd91.114019}
         & NLOG \cite{prd91.114019} \\ \hline
    ${\cal B}r(B^{-}{\to}{\pi}^{-}{\pi}^{0})$
  & $3.5$ & $4.0^{+3.4}_{-1.9}$ & $6.6$
  & $3.54$ & $4.27^{+1.85}_{-1.47}$
  & $3.35^{+1.10}_{-0.80}$ & $4.45^{+1.43}_{-1.05}$\\
    ${\cal B}r(\overline{B}^{0}{\to}{\pi}^{+}{\pi}^{-})$
  & $7.0$ & $6.5^{+6.7}_{-3.8}$ & $6.4$
  & $7.46$ & $7.67^{+3.47}_{-2.64}$
  & $6.19^{+2.12}_{-1.52}$ & $5.39^{+1.88}_{-1.33}$ \\
    ${\cal B}r(\overline{B}^{0}{\to}{\pi}^{0}{\pi}^{0})$
  & $0.12$ & $0.29^{+0.50}_{-0.20}$ & $1.2$
  & $0.12$ & $0.23^{+0.19}_{-0.15}$
  & $0.29^{+0.11}_{-0.07}$ & $0.61^{+0.21}_{-0.17}$ \\ \hline
    $C_{f}(\overline{B}^{0}{\to}{\pi}^{+}{\pi}^{-})$
  & $-14$ & $-18^{+20}_{-12}$ & $-17$
  & $-27$ & $-12^{+4}_{-6}$ \\
    $S_{f}(\overline{B}^{0}{\to}{\pi}^{+}{\pi}^{-})$
  & $-34$ & $-43^{+100}_{-~56}$ & $-43$
  & $-28$ & $-40^{+5}_{-4}$
  \end{tabular}
  \end{ruledtabular}
  \footnotetext[3]{The glauber phases $S_{e}$ $=$ $S_{e1}$ $=$
  $S_{e2}$ $=$ ${\displaystyle -\frac{\pi}{2} }$ is assumed.}
  \end{table}

  (5)
  For the $\overline{B}^{0}$ ${\to}$ $\overline{K}K$ decays,
  the pQCD results with the contributions of WF ${\phi}_{B2}$ are
  in good agreement with data of Belle and BaBar within
  uncertainties. But there are signs of tension between the
  pQCD result with the contributions of WF ${\phi}_{B2}$ and
  experimental data for the $B^{-}$ ${\to}$ $K^{0}K^{-}$ decay.
  In addition, from Table \ref{table-cp-bd}, there is
  a particularly interesting phenomenon that the direct
  $CP$ asymmetries $C_{f}$ is in general larger than the
  mixing-induced $CP$ asymmetries $S_{f}$ for the $B$ ${\to}$
  $PP$ decays, but the opposite is true for the pure
  annihilation $\overline{B}^{0}$ ${\to}$ $K^{+}K^{-}$ decay.

  \begin{figure}[ht]
  \includegraphics[width=0.65\textwidth]{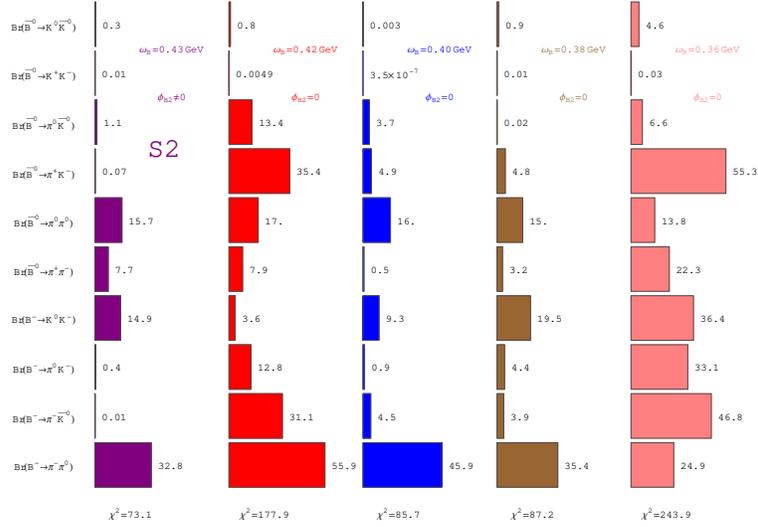}
  \caption{The ${\chi}^{2}$ distribution of branching ratios
  between Belle data and pQCD results with and
  without the participation of ${\phi}_{B2}$
  for different ${\omega}_{B}$, where ${\mu}_{P}$ $=$
  $1.6$ GeV is fixed.}
  \label{br-chi-b1}
  \end{figure}

  (6)
  As pointed out earlier, the formfactors and branching ratios
  are sensitive to the shape parameter ${\omega}_{B}$
  for $B$ mesonic WFs.
  It is noticed that a relatively large value of the parameter
  ${\omega}_{B}$ is optimized for three scenarios by using the
  minimum ${\chi}^{2}$ method.
  An appropriate choice choice of ${\omega}_{B}$ could enhance the
  contributions from ${\phi}_{B1}$ to branching ratios and
  the effects of ${\phi}_{B2}$ could be partly covered, which might
  be one reason why the contributions of ${\phi}_{B2}$ are often
  not considered seriously in many previous studies.
  To illustrate the unique role of ${\phi}_{B2}$,
  the good consistencies between Belle data and pQCD results with
  different ${\omega}_{B}$ are shown in Fig. \ref{br-chi-b1}.
  It is seen that (i) the contributions of ${\phi}_{B2}$ could
  be substituted by an appropriate parameter ${\omega}_{B}$ for
  some cases; (ii) a more comprehensive agreement of branching
  ratios between pQCD calculations and experimental data is
  improved by the participation of ${\phi}_{B2}$.
  In addition, although it is difficult to resolve the $CP$
  puzzles with the LO contributions from ${\phi}_{B2}$, the
  effects of the NLO contributions from ${\phi}_{B2}$ on the
  $CP$ violation might be significant.

  In summary, the $B$ mesonic WF ${\phi}_{B2}$ can contribute
  to emission amplitudes and nonfactorizable annihilation
  amplitudes with the pQCD approach. The enhancements from
  ${\phi}_{B2}$ to hadronic transition formfactors and branching
  ratios for the nonleptonic $B$ ${\to}$ $PP$ decays are
  comparable with those from the NLO corrections without
  taking the ${\phi}_{B2}$ into account.
  By considering ${\phi}_{B2}$, the $B$ ${\to}$ $K{\pi}$ decays
  could be well explained at the LO levels.
  However, the LO contributions from ${\phi}_{B2}$ cannot
  simultaneously settle the branching ratios and $CP$ asymmetries
  for the $\overline{B}^{0}$ ${\to}$ ${\pi}{\pi}$ decays.
  A more careful study of the NLO corrections and other
  effects should be considered in the future.
  In addition, the contributions from ${\phi}_{B2}$ result in
  a small reductions on the $CP$ asymmetries.
  All in all, the oft-ignored ${\phi}_{B2}$ in previous studies
  should be given due attention in order to match the
  improvement of theoretical and experimental results.

  \section*{Acknowledgments}
  The work is supported by the National Natural Science Foundation
  of China (Grant Nos. 11705047, 11981240403, U1632109 and 11547014).

   \begin{appendix}
   \section{The formfactors for $B$ ${\to}$ $P$
    transitions}
   \label{formfactor}
   Besides the definition of Eq.(\ref{formfactor-defination-01}),
   another definition of formfactors is
   \begin{equation}
  {\langle}\,M(p_{2})\,{\vert}\, (\bar{q}\, b)_{V-A}\,
  {\vert} \overline{B}(p_{1})\,{\rangle}\, =\,
    p_{1}^{\mu}\,\tilde{f}_{1}
   +p_{2}^{\mu}\,\tilde{f}_{2}\, =\,
   (p_{1}+p_{2})^{\mu}\,f_{+}
   +q^{\mu}\,f_{-}
   \label{formfactor-defination-02}.
   \end{equation}

   The relations among formfactors are
   \begin{equation}
   F_{1}\, =\, f_{+}\, =\,
   \frac{1}{2}\,\big(\tilde{f}_{1}+\tilde{f}_{2}\big)
   \label{formfactor-defination-03},
   \end{equation}
   \begin{equation}
   F_{0}\, =\,
   \frac{1}{2}\,\tilde{f}_{1}\, \Big( 1+\frac{q^{2}}{m_{B}^{2}} \Big)
  +\frac{1}{2}\,\tilde{f}_{2}\, \Big( 1-\frac{q^{2}}{m_{B}^{2}} \Big)
   \label{formfactor-defination-04}.
   \end{equation}

   Using the pQCDF formula of Eq.(\ref{formfactor-defination-pqcd}),
   the formfactors can be written as follows.
   \begin{equation}
   \tilde{f}_{i}\, =\,
   2\,\frac{{\pi}\,C_{F}}{N_{c}}\,m_{B}^{2}\, f_{B}\,f_{M}\,
    \big( \tilde{f}_{i}^{a}+\tilde{f}_{i}^{b} \big)
   \label{formfactor-defination-05},
   \end{equation}
    \begin{eqnarray}
    \tilde{f}_{1}^{a} &=&
   {\int}_{0}^{1} dx_{1} {\int}_{0}^{1}dx_{2}
   {\int}_{0}^{\infty} db_{1} {\int}_{0}^{\infty} db_{2}\,
   {\alpha}_{s}(t_{a})\,S_{t}(x_{2})\,
    H_{ab}({\alpha}_{g},{\beta}_{a},b_{1},b_{2})
    \nonumber \\ & &
    r_{M}\, e^{-S_{B}}\,e^{-S_{M}}\,
    \Big[ {\phi}_{B1}(x_{1},b_{1})-{\phi}_{B2}(x_{1},b_{1}) \Big]\,
    \Big[ {\phi}_{M}^{p}(x_{2})-{\phi}_{M}^{t}(x_{2}) \Big]
    \label{formfactor-f1-a},
    \end{eqnarray}
    \begin{eqnarray}
    \tilde{f}_{2}^{a} &=&
   {\int}_{0}^{1} dx_{1} {\int}_{0}^{1}dx_{2}
   {\int}_{0}^{\infty} db_{1} {\int}_{0}^{\infty} db_{2}\,
   {\alpha}_{s}(t_{a})\,S_{t}(x_{2})\,
    H_{ab}({\alpha}_{g},{\beta}_{a},b_{1},b_{2})
    \nonumber \\ & &
    e^{-S_{B}}\,e^{-S_{M}}\,
    \Big\{ {\phi}_{B1}(x_{1},b_{1})\,
    \Big[ {\phi}_{M}^{a}(x_{2})\,
    \big(1+x_{2}\,{\eta}\big)
    -2\,r_{M}\,x_{2}\,{\phi}_{M}^{p}(x_{2})
    \nonumber \\ & &
    + 2\,r_{M}\,{\phi}_{M}^{t}(x_{2})\,
    \big( \frac{1}{\eta}-x_{2}\big) \Big]
   -{\phi}_{B2}(x_{1},b_{1})\,
    \Big[ r_{M}\,{\phi}_{M}^{t}(x_{2})\,
    \big( \frac{1}{\eta}-x_{2}\big)
    \nonumber \\ & &
    +{\phi}_{M}^{a}(x_{2})
    -r_{M}\,{\phi}_{M}^{p}(x_{2})\,
    \big( \frac{1}{\eta}+x_{2}\big) \Big] \Big\}
    \label{formfactor-f2-a},
    \end{eqnarray}
    \begin{eqnarray}
    \tilde{f}_{1}^{b} &=&
   {\int}_{0}^{1} dx_{1} {\int}_{0}^{1}dx_{2}
   {\int}_{0}^{\infty} db_{1} {\int}_{0}^{\infty} db_{2}\,
   {\alpha}_{s}(t_{b})\,S_{t}(x_{1})\,
    H_{ab}({\alpha}_{g},{\beta}_{b},b_{2},b_{1})
    \nonumber \\ & &
    e^{-S_{B}}\,e^{-S_{M}}\,
    x_{1}\, \Big\{ {\phi}_{B1}(x_{1},b_{1})\,
    \Big[ {\phi}_{M}^{a}(x_{2})\,{\eta}
    -2\,r_{M}\,{\phi}_{M}^{p}(x_{2}) \Big]
    \nonumber \\ & &
    +{\phi}_{B2}(x_{1},b_{1})\,2\,r_{M}\,
    {\phi}_{M}^{p}(x_{2}) \Big\}
    \label{formfactor-f1-b},
    \end{eqnarray}
    \begin{eqnarray}
    \tilde{f}_{2}^{b} &=&
   {\int}_{0}^{1} dx_{1} {\int}_{0}^{1}dx_{2}
   {\int}_{0}^{\infty} db_{1} {\int}_{0}^{\infty} db_{2}\,
   {\alpha}_{s}(t_{b})\,S_{t}(x_{1})\,
    H_{ab}({\alpha}_{g},{\beta}_{b},b_{2},b_{1})
    \nonumber \\ & &
     e^{-S_{B}}\,e^{-S_{M}}
    \Big\{ {\phi}_{B1}(x_{1},b_{1})\,
    \Big[ 2\,r_{M}\,{\phi}_{M}^{p}(x_{2})\,
    \big(1+\frac{x_{1}}{\eta} \big)
    -x_{1}\,{\phi}_{M}^{a}(x_{2}) \Big]
    \nonumber \\ & &
    -{\phi}_{B2}(x_{1},b_{1})\,
    {\phi}_{M}^{p}(x_{2})\,\frac{2\,r_{M}\,x_{1}}{\eta} \Big\}
    \label{formfactor-f2-b},
    \end{eqnarray}
  where $N_{c}$ $=$ $3$ is the color number. The color factor
  $C_{F}$ $=$ $\displaystyle \frac{N_{c}^{2}-1}{2\,N_{c}}$
  $=$ $\displaystyle \frac{4}{3}$.
  ${\alpha}_{s}$ is the QCD coupling constant.
  $r_{M}$ $=$ ${\displaystyle \frac{{\mu}_{M}}{m_{B}} }$
  and
  ${\eta}$ $=$ $1$ $-$ ${\displaystyle \frac{q^{2}}{m_{B}^{2}} }$.
  The parametrization of $S_{t}(x)$ can be found in
  Ref. \cite{plb555.197}.
  Other parameters are written as follows.
   \begin{equation}
   H_{ab}({\alpha},{\beta},b_{i},b_{j})\, =\,
   b_{i}\,b_{j}\, K_{0}(b_{i}\sqrt{\alpha})\, \Big\{
  {\theta}(b_{i}-b_{j}) K_{0}(b_{i}\sqrt{\beta})\,
   I_{0}(b_{j}\sqrt{\beta})
   +(b_{i} {\leftrightarrow} b_{j}) \Big\}
   \label{amplitude-hab},
   \end{equation}
   \begin{equation}
   S_{B}\, =\, s(x_{1},b_{1},p_{1}^{+})
  +2\,{\int}_{1/b_{1}}^{t}\frac{d{\mu}}{\mu}{\gamma}_{q}
   \label{sudakov-b-meson},
   \end{equation}
   \begin{equation}
   S_{M}\, =\, s(x_{2},b_{2},p_{2}^{-})
   + s(\bar{x}_{2},b_{2},p_{2}^{-})
   +2\,{\int}_{1/b_{2}}^{t}\frac{d{\mu}}{\mu}{\gamma}_{q}
   \label{sudakov-final-meson},
   \end{equation}
   \begin{equation}
  {\alpha}_{g}\, =\, x_{1}\,x_{2}\,{\eta}\,m_{B}^{2}
   \label{alpha-gluon},
   \end{equation}
   \begin{equation}
  {\beta}_{a}\, =\, x_{2}\,{\eta}\,m_{B}^{2}
   \label{beta-a},
   \end{equation}
   \begin{equation}
  {\beta}_{b}\, =\, x_{1}\,{\eta}\,m_{B}^{2}
   \label{beta-b},
   \end{equation}
   \begin{equation}
   t_{a(b)}\, =\, {\max}\big(\sqrt{{\beta}_{a(b)}},
   \frac{1}{b_{1}},\frac{1}{b_{2}}\big)
   \label{scale-tab},
   \end{equation}
  where $I_{0}$ and $K_{0}$ are Bessel functions.
  The expression of $s(x,b,Q)$ can be found in Ref.\cite{prd52.3958}.
  ${\gamma}_{q}$ $=$ ${\displaystyle -\frac{{\alpha}_{s}}{\pi} }$
  is the quark anomalous dimension.
  ${\alpha}_{g}$ and ${\beta}_{a(b)}$ are the virtualities
  of gluon and quarks, respectively.

   \section{The amplitudes for $B$ ${\to}$ ${\pi}{\pi}$, ${\pi}K$,
   $K\overline{K}$ decays}
   \label{decay-modes}
   The amplitude of $B$ meson nonleptonic weak decay is written as
   \begin{equation}
   \mathcal{A}(\overline{B}{\to}M_{1}M_{2})\, =\,
   {\langle}M_{1}M_{2}{\vert}\mathcal{H}_{\rm eff}
   {\vert}\overline{B}{\rangle}
   \label{amplitude-01},
   \end{equation}
   where $\mathcal{H}_{\rm eff}$ is given in Eq.(\ref{hamilton}).

   The explicit expressions for specific final states are written as
   \begin{eqnarray} & &
   \mathcal{A}(B_{u}^{-}{\to}{\pi}^{-}{\pi}^{0})
   \nonumber \\ &=&
   \frac{G_{F}}{2}\, V_{ub}\,V_{ud}^{\ast}\,
   \Big\{ a_{2} \mathcal{A}_{ab}^{LL}({\pi}^{-},{\pi}^{0})
   +C_{1} \mathcal{A}_{cd}^{LL}({\pi}^{-},{\pi}^{0})
   +a_{1} \mathcal{A}_{ab}^{LL}({\pi}^{0},{\pi}^{-})
   +C_{2} \mathcal{A}_{cd}^{LL}({\pi}^{0},{\pi}^{-}) \Big\}
   \nonumber \\ &-&
   \frac{G_{F}}{2}\, V_{tb}\,V_{td}^{\ast}\, \Big\{
   \frac{3}{2}(a_{9}-a_{7}) \mathcal{A}_{ab}^{LL}({\pi}^{-},{\pi}^{0})
  +\frac{1}{2}a_{10} \mathcal{A}_{ab}^{LL}({\pi}^{-},{\pi}^{0})
  +\frac{1}{2}a_{8} \mathcal{A}_{ab}^{SP}({\pi}^{-},{\pi}^{0})
   \nonumber \\ & & 
  +\frac{1}{2}C_{9} \mathcal{A}_{cd}^{LL}({\pi}^{-},{\pi}^{0})
  +\frac{3}{2}C_{10}\mathcal{A}_{cd}^{LL}({\pi}^{-},{\pi}^{0})
  +\frac{3}{2}C_{8}\mathcal{A}_{cd}^{LR}({\pi}^{-},{\pi}^{0})
  +\frac{1}{2}C_{7} \mathcal{A}_{cd}^{SP}({\pi}^{-},{\pi}^{0})
  \nonumber \\ & &
  + a_{10}\mathcal{A}_{ab}^{LL}({\pi}^{0},{\pi}^{-})
  + a_{8}\mathcal{A}_{ab}^{SP}({\pi}^{0},{\pi}^{-})
  + C_{9} \mathcal{A}_{cd}^{LL}({\pi}^{0},{\pi}^{-})
  +C_{7} \mathcal{A}_{cd}^{SP}({\pi}^{0},{\pi}^{-}) \Big\}
   \label{pimpiz-amp},
   \end{eqnarray}
   \begin{eqnarray} & &
   \mathcal{A}(B_{u}^{-}{\to}{\pi}^{-}\overline{K}^{0})
   \nonumber \\ &=&
   \frac{G_{F}}{\sqrt{2}}\, V_{ub}\,V_{us}^{\ast}\,
   \Big\{ a_{1}\, \mathcal{A}_{ef}^{LL}(\overline{K},{\pi})
   +C_{2}\, \mathcal{A}_{gh}^{LL}(\overline{K},{\pi}) \Big\}
   \nonumber \\ &-&
   \frac{G_{F}}{\sqrt{2}}\, V_{tb}\,V_{ts}^{\ast}\,
   \Big\{ (a_{4}-\frac{1}{2}\,a_{10})\,
   \mathcal{A}_{ab}^{LL}({\pi},\overline{K})
   +(a_{6}-\frac{1}{2}\,a_{8})\,
   \mathcal{A}_{ab}^{SP}({\pi},\overline{K})
   \nonumber \\ & & \qquad\ \qquad
   +(C_{3}-\frac{1}{2}\,C_{9})\,
   \mathcal{A}_{cd}^{LL}({\pi},\overline{K})
   +(C_{5}-\frac{1}{2}\,C_{7})\,
   \mathcal{A}_{cd}^{SP}({\pi},\overline{K})
   \nonumber \\ & & \qquad\ \qquad
   +(a_{4}+a_{10})\,\mathcal{A}_{ef}^{LL}(\overline{K},{\pi})
   +(a_{6}+a_{8})\,\mathcal{A}_{ef}^{SP}(\overline{K},{\pi})
   \nonumber \\ & & \qquad\ \qquad
   +(C_{3}+C_{9})\, \mathcal{A}_{gh}^{LL}(\overline{K},{\pi})
   +(C_{5}+C_{7})\,\mathcal{A}_{gh}^{SP}(\overline{K},{\pi}) \Big\}
   \label{pimkz-amp},
   \end{eqnarray}
   \begin{eqnarray} & &
   \mathcal{A}(B_{u}^{-}{\to}K^{-}{\pi}^{0})
   \nonumber \\ &=&
   \frac{G_{F}}{2}\, V_{ub}\,V_{us}^{\ast}\,
   \Big\{ a_{1}\, \mathcal{A}_{ab}^{LL}({\pi},\overline{K})
   +a_{2}\,\mathcal{A}_{ab}^{LL}(\overline{K},{\pi})
   +a_{1}\,\mathcal{A}_{ef}^{LL}(\overline{K},{\pi})
   \nonumber \\ & & \qquad\ \qquad\
   +C_{2} \mathcal{A}_{cd}^{LL}({\pi},\overline{K})
   +C_{1} \mathcal{A}_{cd}^{LL}(\overline{K},{\pi})
   +C_{2} \mathcal{A}_{gh}^{LL}(\overline{K},{\pi}) \Big\}
   \nonumber \\ &-&
   \frac{G_{F}}{2}\, V_{tb}\,V_{ts}^{\ast}\, \Big\{
   ( a_{4}+a_{10} )\, \mathcal{A}_{ab}^{LL}({\pi},\overline{K})
  +( a_{6}+a_{8} )\, \mathcal{A}_{ab}^{SP}({\pi},\overline{K})
   \nonumber \\ & & \qquad\ \qquad
  +( C_{3}+C_{9} )\, \mathcal{A}_{cd}^{LL}({\pi},\overline{K})
  +( C_{5}+C_{7} )\, \mathcal{A}_{cd}^{SP}({\pi},\overline{K})
   \nonumber \\ & & \qquad \qquad
  +\frac{3}{2}\, (a_{9}-a_{7})\, \mathcal{A}_{ab}^{LL}(\overline{K},{\pi})
  +( a_{4}+a_{10} )\, \mathcal{A}_{ef}^{LL}(\overline{K},{\pi})
   \nonumber \\ & & \qquad \qquad
  +( a_{6}+a_{8} )\, \mathcal{A}_{ef}^{SP}(\overline{K},{\pi})
  +\frac{3}{2}\, C_{10}\, \mathcal{A}_{cd}^{LL}(\overline{K},{\pi})
  +\frac{3}{2}\, C_{8}\,  \mathcal{A}_{cd}^{LR}(\overline{K},{\pi})
   \nonumber \\ & & \qquad \qquad
  +( C_{3}+C_{9} )\, \mathcal{A}_{gh}^{LL}(\overline{K},{\pi})
  +( C_{5}+C_{7} )\, \mathcal{A}_{gh}^{SP}(\overline{K},{\pi}) \Big\}
   \label{pizkm-amp},
   \end{eqnarray}
   \begin{eqnarray} & &
   \mathcal{A}(B_{u}^{-}{\to}K^{-}K^{0})
   \nonumber \\ &=&
   \frac{G_{F}}{\sqrt{2}}\, V_{ub}\,V_{ud}^{\ast}\,
   \Big\{ a_{1}\, \mathcal{A}_{ef}^{LL}(K,\overline{K})
   +C_{2}\, \mathcal{A}_{gh}^{LL}(K,\overline{K}) \Big\}
   \nonumber \\ &-&
   \frac{G_{F}}{\sqrt{2}}\, V_{tb}\,V_{td}^{\ast}\, \Big\{
   ( a_{4}-\frac{1}{2}a_{10}) \mathcal{A}_{ab}^{LL}(\overline{K},K)
  +( a_{6}-\frac{1}{2}a_{8} ) \mathcal{A}_{ab}^{SP}(\overline{K},K)
   \nonumber \\ & & \qquad \qquad\
  +( C_{3}-\frac{1}{2}C_{9} ) \mathcal{A}_{cd}^{LL}(\overline{K},K)
  +( C_{5}-\frac{1}{2}C_{7} ) \mathcal{A}_{cd}^{SP}(\overline{K},K)
   \nonumber \\ & & \qquad \qquad\
  +( a_{4}+a_{10}) \mathcal{A}_{ef}^{LL}(K,\overline{K})
  +( a_{6}+a_{8} ) \mathcal{A}_{ef}^{SP}(K,\overline{K})
   \nonumber \\ & & \qquad \qquad\
  +( C_{3}+C_{9} ) \mathcal{A}_{gh}^{LL}(K,\overline{K})
  +( C_{5}+C_{7} ) \mathcal{A}_{gh}^{SP}(K,\overline{K}) \Big\}
   \label{kmkz-amp},
   \end{eqnarray}
   \begin{eqnarray} & &
   \mathcal{A}(\overline{B}_{d}^{0}{\to}{\pi}^{-}{\pi}^{+})\
   \nonumber \\ &=&
   \frac{G_{F}}{\sqrt{2}}\, V_{ub}\,V_{ud}^{\ast}\,
   \Big\{ a_{1} \mathcal{A}_{ab}^{LL}({\pi},{\pi})
   +C_{2} \mathcal{A}_{cd}^{LL}({\pi},{\pi})
   +a_{2} \mathcal{A}_{ef}^{LL}({\pi},{\pi})
   +C_{1} \mathcal{A}_{gh}^{LL}({\pi},{\pi}) \Big\}
   \nonumber \\ &-&
   \frac{G_{F}}{\sqrt{2}}\, V_{tb}\,V_{td}^{\ast}\, \Big\{
    (a_{4}+a_{10})\mathcal{A}_{ab}^{LL}({\pi},{\pi})
   +(a_{6}+a_{8} )\mathcal{A}_{ab}^{SP}({\pi},{\pi})
   \nonumber \\ & & \qquad\ \qquad\
   +( C_{3}+C_{9} ) \mathcal{A}_{cd}^{LL}({\pi},{\pi})
   +( C_{5}+C_{7} ) \mathcal{A}_{cd}^{SP}({\pi},{\pi})
   \nonumber \\ & & \qquad\ \qquad\
  +( 2a_{3}+a_{4}+2a_{5}+\frac{1}{2}a_{7}+\frac{1}{2}a_{9}
  -\frac{1}{2}a_{10}) \mathcal{A}_{ef}^{LL}({\pi},{\pi})
   \nonumber \\ & & \qquad\ \qquad\
  +(a_{6}-\frac{1}{2}a_{8} ) \mathcal{A}_{ef}^{SP}({\pi},{\pi})
  +(C_{3}+2C_{4}-\frac{1}{2}C_{9}+\frac{1}{2}C_{10} )
   \mathcal{A}_{gh}^{LL}({\pi},{\pi})
  \nonumber \\ & & \qquad\ \qquad\
  +(2C_{6}+\frac{1}{2}C_{8}) \mathcal{A}_{gh}^{LR}({\pi},{\pi})
  +(C_{5}-\frac{1}{2}C_{7} ) \mathcal{A}_{gh}^{SP}({\pi},{\pi}) \Big\}
   \label{pimpip-amp},
   \end{eqnarray}
   \begin{eqnarray} & &
   \mathcal{A}(\overline{B}_{d}^{0}{\to}{\pi}^{0}{\pi}^{0})
   \nonumber \\ &=&
   -\frac{G_{F}}{\sqrt{2}}\, V_{ub}\,V_{ud}^{\ast}\,
   \Big\{ a_{2} \big(\mathcal{A}_{ab}^{LL}-\mathcal{A}_{ef}^{LL} \big)
   +C_{1} \big(\mathcal{A}_{cd}^{LL}- \mathcal{A}_{gh}^{LL} \big) \Big\}
   \nonumber \\ & &
  -\frac{G_{F}}{\sqrt{2}}\, V_{tb}\,V_{td}^{\ast}\,
   \Big\{ (a_{4}+\frac{3}{2}a_{7}-\frac{3}{2}a_{9}-
   \frac{1}{2}a_{10}) \mathcal{A}_{ab}^{LL}
  +(a_{6}-\frac{1}{2}a_{8})
   \big( \mathcal{A}_{ab}^{SP}+\mathcal{A}_{ef}^{SP} \big)
   \nonumber \\ & & \qquad
   +(C_{3}-\frac{1}{2}C_{9}-\frac{3}{2}C_{10}) \mathcal{A}_{cd}^{LL}
   -\frac{3}{2} C_{8} \mathcal{A}_{cd}^{LR}
   +(C_{5}-\frac{1}{2}C_{7}) \big( \mathcal{A}_{cd}^{SP}
   +\mathcal{A}_{gh}^{SP} \big)
   \nonumber \\ & & \qquad
   +(2a_{3}+a_{4}+2a_{5}+\frac{1}{2}a_{7}+\frac{1}{2}a_{9}-
   \frac{1}{2}a_{10}) \mathcal{A}_{ef}^{LL}
  +(2C_{6}+\frac{1}{2}C_{8}) \mathcal{A}_{gh}^{LR}
   \nonumber \\ & & \qquad
  +(C_{3}+2C_{4}-\frac{1}{2}C_{9}+\frac{1}{2}C_{10})
   \mathcal{A}_{gh}^{LL} \Big\}
   \label{pizpiz-amp},
   \end{eqnarray}
   \begin{eqnarray} & &
   \mathcal{A}(\overline{B}_{d}^{0}{\to}K^{-}{\pi}^{+})
   \nonumber \\ &=&
   \frac{G_{F}}{\sqrt{2}}\, V_{ub}\,V_{us}^{\ast}\,
   \Big\{ a_{1} \mathcal{A}_{ab}^{LL}({\pi},\overline{K})
  +C_{2}\mathcal{A}_{cd}^{LL}({\pi},\overline{K}) \Big\}
   \nonumber \\ &-&
   \frac{G_{F}}{\sqrt{2}}\, V_{tb}\,V_{ts}^{\ast}\, \Big\{
   (a_{4}+a_{10}) \mathcal{A}_{ab}^{LL}({\pi},\overline{K})
  +(a_{6}+a_{8}) \mathcal{A}_{ab}^{SP}({\pi},\overline{K})
  \nonumber \\ & & \qquad \qquad\
  +(C_{3}+C_{9})\mathcal{A}_{cd}^{LL}({\pi},\overline{K})
  +(C_{5}+C_{7})\mathcal{A}_{cd}^{SP}({\pi},\overline{K})
   \nonumber \\ & & \qquad \qquad\
  +(a_{4}-\frac{1}{2}a_{10}) \mathcal{A}_{ef}^{LL}(\overline{K},{\pi})
  +(a_{6}-\frac{1}{2}a_{8} ) \mathcal{A}_{ef}^{SP}(\overline{K},{\pi})
   \nonumber \\ & & \qquad \qquad\
  +(C_{3}-\frac{1}{2}C_{9}) \mathcal{A}_{gh}^{LL}(\overline{K},{\pi})
  +(C_{5}-\frac{1}{2}C_{7} ) \mathcal{A}_{gh}^{SP}(\overline{K},{\pi}) \Big\}
   \label{pipkm-amp},
   \end{eqnarray}
   \begin{eqnarray} & &
   \mathcal{A}(\overline{B}_{d}^{0}{\to}\overline{K}^{0}{\pi}^{0})
   \nonumber \\ &=&
   \frac{G_{F}}{2}\, V_{ub}\,V_{us}^{\ast}\,
   \Big\{ a_{2} \mathcal{A}_{ab}^{LL}(\overline{K},{\pi})
   + C_{1} \mathcal{A}_{cd}^{LL}(\overline{K},{\pi}) \Big\}
   \nonumber \\ &+&
   \frac{G_{F}}{2}\, V_{tb}\,V_{ts}^{\ast}\, \Big\{
   (a_{4}-\frac{1}{2}a_{10})
   \mathcal{A}_{ab}^{LL}({\pi},\overline{K})
  +(a_{6}-\frac{1}{2}a_{8})
   \mathcal{A}_{ab}^{SP}({\pi},\overline{K})
   \nonumber \\ & & \qquad \qquad\
  +\frac{3}{2} (a_{7}-a_{9})
   \mathcal{A}_{ab}^{LL}(\overline{K},{\pi})
  +(a_{4}-\frac{1}{2}a_{10})
   \mathcal{A}_{ef}^{LL}(\overline{K},{\pi})
   \nonumber \\ & & \qquad \qquad\
  +(a_{6}-\frac{1}{2}a_{8})
   \mathcal{A}_{ef}^{SP}(\overline{K},{\pi})
  +(C_{3}-\frac{1}{2}C_{9})
   \mathcal{A}_{cd}^{LL}({\pi},\overline{K})
   \nonumber \\ & & \qquad \qquad\
  +(C_{5}-\frac{1}{2}C_{7})
   \mathcal{A}_{cd}^{SP}({\pi},\overline{K})
  -\frac{3}{2}C_{8}\mathcal{A}_{cd}^{LR}(\overline{K},{\pi})
  -\frac{3}{2}C_{10}\mathcal{A}_{cd}^{LL}(\overline{K},{\pi})
   \nonumber \\ & & \qquad \qquad\
  +(C_{3}-\frac{1}{2}C_{9})
   \mathcal{A}_{gh}^{LL}(\overline{K},{\pi})
  +(C_{5}-\frac{1}{2}a_{7})
   \mathcal{A}_{gh}^{SP}(\overline{K},{\pi}) \Big\}
   \label{pizkz-amp},
   \end{eqnarray}
   \begin{eqnarray} & &
   \mathcal{A}(\overline{B}_{d}^{0}{\to}K^{-}K^{+})
   \nonumber \\ &=&
   \frac{G_{F}}{\sqrt{2}}\, V_{ub}\,V_{ud}^{\ast}\,
   \Big\{ a_{2} \mathcal{A}_{ef}^{LL}(K,\overline{K})
  +C_{1} \mathcal{A}_{gh}^{LL}(K,\overline{K}) \Big\}
   \nonumber \\ &-&
   \frac{G_{F}}{\sqrt{2}}\, V_{tb}\,V_{td}^{\ast}\, \Big\{
   ( a_{3}+a_{5}+a_{7}+a_{9} ) \mathcal{A}_{ef}^{LL}(K,\overline{K})
   \nonumber \\ & & \qquad\ \qquad
   +( a_{3}+a_{5}-\frac{1}{2}a_{7}-\frac{1}{2}a_{9} )
   \mathcal{A}_{ef}^{LL}(\overline{K},K)
   \nonumber \\ & & \qquad\ \qquad
   +( C_{4}+C_{10} )\mathcal{A}_{gh}^{LL}(K,\overline{K})
   +( C_{6}+C_{8} ) \mathcal{A}_{gh}^{LR}(K,\overline{K})
   \nonumber \\ & & \qquad\ \qquad
   +( C_{4}-\frac{1}{2}C_{10} )
   \mathcal{A}_{gh}^{LL}(\overline{K},K)
   +( C_{6}-\frac{1}{2}C_{8} )
   \mathcal{A}_{gh}^{LR}(\overline{K},K) \Big\}
   \label{kmkp-amp},
   \end{eqnarray}
   \begin{eqnarray} & &
   \mathcal{A}(\overline{B}_{d}^{0}{\to}\overline{K}^{0}K^{0})
   \nonumber \\ &=& -
   \frac{G_{F}}{\sqrt{2}}\, V_{tb}\,V_{td}^{\ast}\, \Big\{
   (a_{4}-\frac{1}{2}a_{10}) \mathcal{A}_{ab}^{LL}(\overline{K},K)
  +(a_{6}-\frac{1}{2}a_{8}) \mathcal{A}_{ab}^{SP}(\overline{K},K)
   \nonumber \\ & & \qquad \qquad \quad\
  +(C_{3}-\frac{1}{2}C_{9}) \mathcal{A}_{cd}^{LL}(\overline{K},K)
  +(C_{5}-\frac{1}{2}C_{7}) \mathcal{A}_{cd}^{SP}(\overline{K},K)
   \nonumber \\ & & \qquad \qquad \quad\
  +(a_{3}+a_{5}-\frac{1}{2}a_{7}-\frac{1}{2}a_{9})
   \mathcal{A}_{ef}^{LL}(\overline{K},K)
  +(C_{4}-\frac{1}{2}C_{10}) \mathcal{A}_{gh}^{LL}(\overline{K},K)
   \nonumber \\ & & \qquad \qquad \quad\
  +(C_{6}-\frac{1}{2}C_{8}) \mathcal{A}_{gh}^{LR}(\overline{K},K)
  +(a_{6}-\frac{1}{2}a_{8}) \mathcal{A}_{ef}^{SP}(K,\overline{K})
   \nonumber \\ & & \qquad \qquad \quad\
   +(a_{3}+a_{4}+a_{5}-\frac{1}{2}a_{7}-\frac{1}{2}a_{9}
   -\frac{1}{2}a_{10}) \mathcal{A}_{ef}^{LL}(K,\overline{K})
   \nonumber \\ & & \qquad \qquad \quad\
   +(C_{3}+C_{4}-\frac{1}{2}C_{9}-\frac{1}{2}C_{10})
   \mathcal{A}_{gh}^{LL}(K,\overline{K})
   \nonumber \\ & & \qquad \qquad \quad\
   +(C_{6}-\frac{1}{2}C_{8})
   \mathcal{A}_{gh}^{LR}(K,\overline{K})
   +(C_{5}-\frac{1}{2}C_{7})
   \mathcal{A}_{gh}^{SP}(K,\overline{K}) \Big\}
   \label{kzkz-amp}.
   \end{eqnarray}
   The shorthands are
   \begin{equation}
   a_{i}\, =\, \left\{ \begin{array}{lll}
   \displaystyle C_{i}+\frac{1}{N_{c}}C_{i+1},
   & \quad & \text{for odd }i; \\
   \displaystyle C_{i}+\frac{1}{N_{c}}C_{i-1},
   & \quad & \text{for even }i,
   \end{array} \right.
   \label{wilson-ai}
   \end{equation}
   \begin{equation}
   C_{m}\,\mathcal{A}_{ij}^{k}(M_{1},M_{2})\, =\,
   i\,\mathcal{K}\,\frac{{\pi}\,C_{F}}{N_{c}^{2}}\, m_{B}^{4}\,
   f_{B}\,f_{M_{1}}\,f_{M_{2}}\, \big\{
   \mathcal{A}_{i}^{k}(C_{m},M_{1},M_{2})
  +\mathcal{A}_{j}^{k}(C_{m},M_{1},M_{2}) \big\}
   \label{mode-factor-01},
   \end{equation}
   where the factor $\mathcal{K}$ $=$ $N_{c}$ for factorizable amplitudes
   with $ij$ $=$ $ab$ and $ef$, and $\mathcal{K}$ $=$ $1$ for
   nonfactorizable amplitudes with $ij$ $=$ $cd$ and $gh$.
   The expressions of amplitude building block
   $\mathcal{A}_{i}^{k}(C_{m},M_{1},M_{2})$ are given in
   Appendix \ref{blocks}.

  \section{Amplitude building blocks}
  \label{blocks}
  There should be a Sudakov factor corresponding to each WF with
  the pQCD approach.
  For the sake of convenience in writing, the shorthands such as
  ${\phi}_{B1}$ $=$ ${\phi}_{B1}(x_{1},b_{1})\,e^{-S_{B}}$,
  ${\phi}_{B2}$ $=$ ${\phi}_{B2}(x_{1},b_{1})\,e^{-S_{B}}$,
  ${\phi}_{M}^{a}$ $=$ ${\phi}_{M}^{a}(x_{2})\,e^{-S_{M}}$,
  ${\phi}_{M^{\prime}}^{a}$ $=$
  ${\phi}_{M^{\prime}}^{a}(x_{3})\,e^{-S_{M^{\prime}}}$,
  ${\phi}_{M}^{p,t}$ $=$ $r_{M}\,{\phi}_{M}^{p,t}(x_{2})\,e^{-S_{M}}$,
  ${\phi}_{M^{\prime}}^{p,t}$ $=$
  $r_{M^{\prime}}\,{\phi}_{M^{\prime}}^{p,t}(x_{3})\,e^{-S_{M^{\prime}}}$
  and
  $\mathcal{A}_{i}^{j}$ $=$ $\mathcal{A}_{i}^{j}(C_{m},M,M^{\prime})$
  will be used in this section.
  As to the amplitude building block $\mathcal{A}_{i}^{j}$,
  the subscript $i$ corresponds to the indices of
  Fig.\ref{fig-amplitude-01}, and the superscript
  $j$ refers to the three possible Dirac structures
  ${\Gamma}_{1}{\otimes}{\Gamma}_{2}$ of the operator
  $(\bar{q}_{1}q_{2})_{{\Gamma}_{1}}(\bar{q}_{3}q_{4})_{{\Gamma}_{2}}$,
  namely $j$ $=$ $LL$ for $(V-A){\otimes}(V-A)$, $j$ $=$ $LR$
  for $(V-A){\otimes}(V+A)$ and $j$ $=$ $SP$ for $
  -2\,(S-P){\otimes}(S+P)$.
  The expressions of $\mathcal{A}_{i}^{j}$ are written as follows.
   \begin{eqnarray}
   \mathcal{A}_{a}^{LL} &=&
  {\int} dx_{1}\,dx_{2}\,db_{1}\, db_{2}\,
  C_{i}(t_{a})\, {\alpha}_{s}(t_{a})\, S_{t}(x_{2})\,
  H_{ab}({\alpha}_{g},{\beta}_{a},b_{1},b_{2})
   \nonumber \\ & &
   \big\{ {\phi}_{B1}\, \big[ {\phi}_{M}^{a}\,
   \big(1+x_{2}\big) + \big( {\phi}_{M}^{p}+ {\phi}_{M}^{t} \big)\,
   \big( \bar{x}_{2}-x_{2}\big) \big]
   \nonumber \\ & &
 -{\phi}_{B2}\, \big[ {\phi}_{M}^{a} -
   \big( {\phi}_{M}^{p}+{\phi}_{M}^{t} \big)\,x_{2} \big] \big\}
   \label{amp-e01-01},
   \end{eqnarray}
   \begin{equation}
   \mathcal{A}_{a}^{LR}\, =\, -\mathcal{A}_{a}^{LL}
   \label{amp-e01-02},
   \end{equation}
   \begin{eqnarray}
   \mathcal{A}_{a}^{SP} &=&
   2\ r_{M^{\prime}}\,
  {\int} dx_{1}\,dx_{2}\,db_{1}\, db_{2}\,
  C_{i}(t_{a})\, {\alpha}_{s}(t_{a})\, S_{t}(x_{2})\,
  H_{ab}({\alpha}_{g},{\beta}_{a},b_{1},b_{2})
   \nonumber \\ & &
   \big\{ {\phi}_{B1}\, \big[ {\phi}_{M}^{a}
   +{\phi}_{M}^{p}\, \big( 2+x_{2}\big)
   -{\phi}_{M}^{t}\, x_{2} \big]
   -{\phi}_{B2}\, \big[ {\phi}_{M}^{a}+{\phi}_{M}^{p}
   -{\phi}_{M}^{t} \big] \big\}
   \label{amp-e01-03},
   \end{eqnarray}
   \begin{equation}
   \mathcal{A}_{b}^{LL}\, =\, 2\,
  {\int} dx_{1}\,dx_{2}\,db_{1}\, db_{2}\,
  C_{i}(t_{b})\, {\alpha}_{s}(t_{b})\,S_{t}(x_{1})\,
  H_{ab}({\alpha}_{g},{\beta}_{b},b_{2},b_{1})\,
  {\phi}_{B1}\,{\phi}_{M}^{p}
   \label{amp-e02-01},
   \end{equation}
   \begin{equation}
   \mathcal{A}_{b}^{LR}\, =\, -\mathcal{A}_{b}^{LL}
   \label{amp-e02-02},
   \end{equation}
   \begin{eqnarray}
   \mathcal{A}_{b}^{SP} &=&
  {\int} dx_{1}\,dx_{2}\,db_{1}\, db_{2}\,
  C_{i}(t_{b})\, {\alpha}_{s}(t_{b})\,S_{t}(x_{1})\,
  H_{ab}({\alpha}_{g},{\beta}_{b},b_{2},b_{1})
   \nonumber \\ & & 2\,r_{M^{\prime}}\,
   \big\{ {\phi}_{B1}\, \big[ {\phi}_{M}^{a}\,x_{1}
   +2\,{\phi}_{M}^{p}\, \bar{x}_{1} \big]
   +2\,{\phi}_{B2}\,{\phi}_{M}^{p}\,x_{1} \big\}
   \label{amp-e02-03},
   \end{eqnarray}
   \begin{eqnarray}
   \mathcal{A}_{c}^{LL} &=&
  {\int} dx_{1}\,dx_{2}\,dx_{3}\,db_{2}\,db_{3}\,
   C_{i}(t_{c})\, {\alpha}_{s}(t_{c})\, S_{t}(x_{2})\,
   H_{cd}({\alpha}_{g},{\beta}_{c},b_{2},b_{3})\,
  {\phi}_{M^{\prime}}^{a}
   \nonumber \\ & &
   \big\{ \big({\phi}_{B1}-{\phi}_{B2}\big)\,{\phi}_{M}^{a}\,
   \big(\bar{x}_{3}-x_{1}\big)-{\phi}_{B1}\,
   \big({\phi}_{M}^{p}-{\phi}_{M}^{t}\big)\,x_{2} \big\}_{b_{1}=b_{2}}
   \label{amp-e03-01},
   \end{eqnarray}
   \begin{eqnarray}
   \mathcal{A}_{c}^{LR} &=&
  {\int} dx_{1}\,dx_{2}\,dx_{3}\,db_{2}\,db_{3}\,
   C_{i}(t_{c})\, {\alpha}_{s}(t_{c})\, S_{t}(x_{2})\,
   H_{cd}({\alpha}_{g},{\beta}_{c},b_{2},b_{3})\,
  {\phi}_{M^{\prime}}^{a}
   \nonumber \\ & & \hspace{-0.05\textwidth}
   \big\{ \big({\phi}_{B1}-{\phi}_{B2} \big)
   \big[ {\phi}_{M}^{a}\,\big( x_{1}-\bar{x}_{3} \big)
  +\big({\phi}_{M}^{p}+{\phi}_{M}^{t}\big)\,x_{2} \big]
  -{\phi}_{B1}\,{\phi}_{M}^{a}\,x_{2} \big\}_{b_{1}=b_{2}}
   \label{amp-e03-02},
   \end{eqnarray}
   \begin{eqnarray}
   \mathcal{A}_{c}^{SP} &=&
  {\int} dx_{1}\,dx_{2}\,dx_{3}\,db_{2}\,db_{3}\,
   C_{i}(t_{c})\, {\alpha}_{s}(t_{c})\, S_{t}(x_{2})\,
   H_{cd}({\alpha}_{g},{\beta}_{c},b_{2},b_{3})
   \nonumber \\ & &
   \big\{ \big( {\phi}_{B1}-{\phi}_{B2} \big)
   \big( {\phi}_{M}^{a}+ {\phi}_{M}^{p}-{\phi}_{M}^{t} \big)
   \big( {\phi}_{M^{\prime}}^{p}+{\phi}_{M^{\prime}}^{t} \big)
   \big(\bar{x}_{3}-x_{1}\big)
   \nonumber \\ & & +\,
  {\phi}_{B1}\,x_{2}\,\big({\phi}_{M}^{p}+{\phi}_{M}^{t}\big)
   \big({\phi}_{M^{\prime}}^{p}-{\phi}_{M^{\prime}}^{t} \big)
   \big\}_{b_{1}=b_{2}}
   \label{amp-e03-03},
   \end{eqnarray}
   \begin{eqnarray}
   \mathcal{A}_{d}^{LL} &=&
  {\int} dx_{1}\,dx_{2}\,dx_{3}\,db_{2}\,db_{3}\,
   C_{i}(t_{d})\, {\alpha}_{s}(t_{d})\, S_{t}(x_{2})\,
   H_{cd}({\alpha}_{g},{\beta}_{d},b_{2},b_{3})\,
   {\phi}_{M^{\prime}}^{a}
    \nonumber \\ & & \hspace{-0.05\textwidth}
    \big\{ \big({\phi}_{B1}-{\phi}_{B2} \big)
    \big[ {\phi}_{M}^{a}\,\big( x_{1}-x_{3}\big)
   + \big({\phi}_{M}^{p}+{\phi}_{M}^{t}\big)\,x_{2} \big]
   -{\phi}_{B1}\,{\phi}_{M}^{a}\,x_{2} \big\}_{b_{1}=b_{2}}
   \label{amp-e04-01},
   \end{eqnarray}
   \begin{eqnarray}
   \mathcal{A}_{d}^{LR} &=&
  {\int} dx_{1}\,dx_{2}\,dx_{3}\,db_{2}\,db_{3}\,
   C_{i}(t_{d})\, {\alpha}_{s}(t_{d})\, S_{t}(x_{2})\,
   H_{cd}({\alpha}_{g},{\beta}_{d},b_{2},b_{3})\,
   {\phi}_{M^{\prime}}^{a}
   \nonumber \\ & &
   \big\{ \big({\phi}_{B1}-{\phi}_{B2} \big)\,{\phi}_{M}^{a}\,
   \big(x_{3}-x_{1}\big) -{\phi}_{B1}\,\big({\phi}_{M}^{p}-
   {\phi}_{M}^{t}\big)\,x_{2} \big\}_{b_{1}=b_{2}}
   \label{amp-e04-02},
   \end{eqnarray}
   \begin{eqnarray}
   \mathcal{A}_{d}^{SP} &=&
  {\int} dx_{1}\,dx_{2}\,dx_{3}\,db_{2}\,db_{3}\,
   C_{i}(t_{d})\, {\alpha}_{s}(t_{d})\, S_{t}(x_{2})\,
   H_{cd}({\alpha}_{g},{\beta}_{d},b_{2},b_{3})
   \nonumber \\ & &
   \big\{ \big({\phi}_{B1}-{\phi}_{B2}\big)\,
   \big( {\phi}_{M}^{a}+{\phi}_{M}^{p}- {\phi}_{M}^{t}\big)\,
   \big({\phi}_{M^{\prime}}^{p}-{\phi}_{M^{\prime}}^{t} \big)\,
   \big( x_{1}-x_{3} \big)
   \nonumber \\ & & -\,
  {\phi}_{B1}\,x_{2}\,\big({\phi}_{M}^{p}+{\phi}_{M}^{t} \big)\,
   \big({\phi}_{M^{\prime}}^{p}+{\phi}_{M^{\prime}}^{t} \big)
   \big\}_{b_{1}=b_{2}}
   \label{amp-e04-03},
   \end{eqnarray}
   \begin{eqnarray}
   \mathcal{A}_{e}^{LL} &=& -
  {\int} dx_{2}\,dx_{3}\,db_{2}\,db_{3}\,
   C_{i}(t_{e})\, {\alpha}_{s}(t_{e})\, S_{t}(\bar{x}_{3})\,
   H_{ef}({\omega}_{g},{\beta}_{e},b_{2},b_{3})
   \nonumber \\ & & \quad
   \big\{ {\phi}_{M}^{a}\,{\phi}_{M^{\prime}}^{a}\, \bar{x}_{3}
   +2\,{\phi}_{M}^{p}\, \big[ {\phi}_{M^{\prime}}^{p}\,
   \big(1+\bar{x}_{3}\big)+{\phi}_{M^{\prime}}^{t}\,x_{3} \big] \big\}
   \label{amp-a01-01},
   \end{eqnarray}
   \begin{equation}
   \mathcal{A}_{e}^{LR}\, =\, +\mathcal{A}_{e}^{LL}
   \label{amp-a01-02},
   \end{equation}
   \begin{eqnarray}
   \mathcal{A}_{e}^{SP} &=&
  {\int} dx_{2}\,dx_{3}\,db_{2}\,db_{3}\,
   C_{i}(t_{e})\, {\alpha}_{s}(t_{e})\, S_{t}(\bar{x}_{3})\,
   H_{ef}({\omega}_{g},{\beta}_{e},b_{2},b_{3})
   \nonumber \\ & & 2\,
   \big\{ {\phi}_{M}^{a}\,\bar{x}_{3}\,
   \big( {\phi}_{M^{\prime}}^{p}+{\phi}_{M^{\prime}}^{t} \big)
  +2\,{\phi}_{M}^{p}\,{\phi}_{M^{\prime}}^{a} \big\}
   \label{amp-a01-03},
   \end{eqnarray}
   \begin{eqnarray}
   \mathcal{A}_{f}^{LL} &=&
  {\int} dx_{2}\,dx_{3}\,db_{2}\,db_{3}\,
   C_{i}(t_{f})\, {\alpha}_{s}(t_{f})\, S_{t}(x_{2})\,
   H_{ef}({\omega}_{g},{\beta}_{f},b_{3},b_{2})
   \nonumber \\ & &
   \big\{ {\phi}_{M}^{a}\,{\phi}_{M^{\prime}}^{a}\,x_{2}
   +2\,\big[ {\phi}_{M}^{p}\,\big(1+x_{2}\big)
   -{\phi}_{M}^{t}\,\bar{x}_{2} \big] {\phi}_{M^{\prime}}^{p} \big\}
   \label{amp-a02-01},
   \end{eqnarray}
   \begin{equation}
   \mathcal{A}_{f}^{LR}\, =\,  +\mathcal{A}_{f}^{LL}
   \label{amp-a02-02},
   \end{equation}
   \begin{eqnarray}
   \mathcal{A}_{f}^{SP} &=&
  {\int} dx_{2}\,dx_{3}\,db_{2}\,db_{3}\,
   C_{i}(t_{f})\, {\alpha}_{s}(t_{f})\, S_{t}(x_{2})\,
   H_{ef}({\omega}_{g},{\beta}_{f},b_{3},b_{2})
   \nonumber \\ & & 2\,
   \big\{ 2\,{\phi}_{M}^{a}\,{\phi}_{M^{\prime}}^{p}
   + x_{2}\,\big({\phi}_{M}^{p}-{\phi}_{M}^{t} \big)\,
   {\phi}_{M^{\prime}}^{a} \big\}
   \label{amp-a02-03},
   \end{eqnarray}
   \begin{eqnarray}
   \mathcal{A}_{g}^{LL} &=&
  {\int} dx_{1}\,dx_{2}\,dx_{3}\,db_{1}\,db_{2}\,
   C_{i}(t_{g})\, {\alpha}_{s}(t_{g})\,
   H_{gh}({\omega}_{g},{\beta}_{g},b_{1},b_{2})
   \nonumber \\ & &
   \big\{ {\phi}_{B2}\, \big( {\phi}_{M}^{p}-{\phi}_{M}^{t}\big)\,
   \big[ {\phi}_{M^{\prime}}^{p}\, \big(x_{1}+\bar{x}_{3}+1\big)+
  {\phi}_{M^{\prime}}^{t}\, \big(x_{1}-x_{3}\big) \big]
   \nonumber \\ & & -\,
   {\phi}_{B1}\, \big[ {\phi}_{M}^{a}\,{\phi}_{M^{\prime}}^{a}\,x_{2}
  +2\,{\phi}_{M}^{p}\, {\phi}_{M^{\prime}}^{p}\,\big(1+x_{2}\big)
  +2\,{\phi}_{M}^{t}\, {\phi}_{M^{\prime}}^{t}\, \bar{x}_{2}
   \nonumber \\ & & \qquad
  +\big( {\phi}_{M}^{p}-{\phi}_{M}^{t}\big)\,
   \big( {\phi}_{M^{\prime}}^{p}+{\phi}_{M^{\prime}}^{t}\big)\,
   \big( x_{1}-x_{2}+\bar{x}_{3}\big) \big] \big\}_{b_{2}=b_{3}}
   \label{amp-a03-01},
   \end{eqnarray}
   \begin{eqnarray}
   \mathcal{A}_{g}^{LR} &=& -
  {\int} dx_{1}\,dx_{2}\,dx_{3}\,db_{1}\,db_{2}\,
   C_{i}(t_{g})\, {\alpha}_{s}(t_{g})\,
   H_{gh}({\omega}_{g},{\beta}_{g},b_{1},b_{2})
   \nonumber \\ & &
   \big\{ \big( {\phi}_{B1}-{\phi}_{B2} \big)\,
  {\phi}_{M}^{a}\,{\phi}_{M^{\prime}}^{a}\,
   \big( x_{1}+\bar{x}_{3} \big)
 +{\phi}_{B1}\, \big[2\, {\phi}_{M}^{p}\,
  {\phi}_{M^{\prime}}^{p}\, \big(1+x_{2}\big)
   \nonumber \\ & & +\,
   2\,{\phi}_{M}^{t}\,{\phi}_{M^{\prime}}^{t}\, \bar{x}_{2}
  +\big( {\phi}_{M}^{p}+{\phi}_{M}^{t}\big)\,
   \big( {\phi}_{M^{\prime}}^{p}-{\phi}_{M^{\prime}}^{t}
   \big)\,\big( x_{1}-x_{2}+\bar{x}_{3}\big)\big]
   \nonumber \\ & & -
  {\phi}_{B2}\, \big( {\phi}_{M^{\prime}}^{p}-
  {\phi}_{M^{\prime}}^{t}\big)\, \big[ {\phi}_{M}^{p}\,
   \big(x_{1}+\bar{x}_{3}+1\big) + {\phi}_{M}^{t}\,
   \big(x_{1}-x_{3}\big) \big] \big\}_{b_{2}=b_{3}}
   \label{amp-a03-02},
   \end{eqnarray}
   \begin{eqnarray}
   \mathcal{A}_{g}^{SP} &=&
  {\int} dx_{1}\,dx_{2}\,dx_{3}\,db_{1}\,db_{2}\,
   C_{i}(t_{g})\, {\alpha}_{s}(t_{g})\,
   H_{gh}({\omega}_{g},{\beta}_{g},b_{1},b_{2})
   \nonumber \\ & & 
   \big\{ {\phi}_{B1}\, \big[ {\phi}_{M}^{a}\,
   \big( {\phi}_{M^{\prime}}^{p}- {\phi}_{M^{\prime}}^{t}\big)\,
   \big( x_{1}+\bar{x}_{3}-2\big)
   + {\phi}_{M^{\prime}}^{a}\, \big( {\phi}_{M}^{p}
   +{\phi}_{M}^{t} \big)\, \big( 2-x_{2}\big) \big]
   \nonumber \\ & & -
  {\phi}_{B2}\,\big[ {\phi}_{M}^{a}\, \big( {\phi}_{M^{\prime}}^{p}-
  {\phi}_{M^{\prime}}^{t}\big)\, \big( x_{1}-x_{3}\big)
   + {\phi}_{M^{\prime}}^{a}\, \big( {\phi}_{M}^{p}
  +{\phi}_{M}^{t} \big) \big] \big\}_{b_{2}=b_{3}}
   \label{amp-a03-03},
   \end{eqnarray}
   \begin{eqnarray}
   \mathcal{A}_{h}^{LL} &=&
  {\int} dx_{1}\,dx_{2}\,dx_{3}\,db_{1}\,db_{2}\,
   C_{i}(t_{h})\, {\alpha}_{s}(t_{h})\,
   H_{gh}({\omega}_{g},{\beta}_{h},b_{1},b_{2})
   \nonumber \\ & &
   \big\{ \big( {\phi}_{B1}-{\phi}_{B2} \big)
   \big( \bar{x}_{3}-x_{1} \big) \big[
  {\phi}_{M}^{a}\,{\phi}_{M^{\prime}}^{a}
  +\big({\phi}_{M}^{p}+{\phi}_{M}^{t} \big)\,
   \big( {\phi}_{M^{\prime}}^{p}-{\phi}_{M^{\prime}}^{t} \big) \big]
   \nonumber \\ & &
  +{\phi}_{B1}\, x_{2}\, \big({\phi}_{M}^{p}-{\phi}_{M}^{t}\big)\,
   \big({\phi}_{M^{\prime}}^{p} +{\phi}_{M^{\prime}}^{t} \big)
   \big\}_{b_{2}=b_{3}}
   \label{amp-a04-01},
   \end{eqnarray}
   \begin{eqnarray}
   \mathcal{A}_{h}^{LR} &=&
  {\int} dx_{1}\,dx_{2}\,dx_{3}\,db_{1}\,db_{2}\,
   C_{i}(t_{h})\, {\alpha}_{s}(t_{h})\,
   H_{gh}({\omega}_{g},{\beta}_{h},b_{1},b_{2})
   \nonumber \\ & &
   \big\{ \big( {\phi}_{B1}-{\phi}_{B2} \big)\,
   \big(\bar{x}_{3}-x_{1}\big)\,
   \big({\phi}_{M}^{p}-{\phi}_{M}^{t}\big)\,
   \big({\phi}_{M^{\prime}}^{p}+{\phi}_{M^{\prime}}^{t}\big)
   \nonumber \\ & & +\,
  {\phi}_{B1}\,x_{2}\, \big[ {\phi}_{M}^{a}\,{\phi}_{M^{\prime}}^{a}
   + \big({\phi}_{M}^{p}+ {\phi}_{M}^{t}\big)\,
   \big( {\phi}_{M^{\prime}}^{p}-{\phi}_{M^{\prime}}^{t} \big)
   \big] \big\}_{b_{2}=b_{3}}
   \label{amp-a04-02},
   \end{eqnarray}
   \begin{eqnarray}
   \mathcal{A}_{h}^{SP} &=&
  {\int} dx_{1}\,dx_{2}\,dx_{3}\,db_{1}\,db_{2}\,
   C_{i}(t_{h})\, {\alpha}_{s}(t_{h})\,
   H_{gh}({\omega}_{g},{\beta}_{h},b_{1},b_{2})
   \nonumber \\ & &
   \big\{ {\phi}_{B1}\, \big[ x_{2}\, {\phi}_{M^{\prime}}^{a}\,
   \big({\phi}_{M}^{p}+{\phi}_{M}^{t}\big)
   + \big(x_{1}-\bar{x}_{3}\big)\,
   {\phi}_{M}^{a}\,\big( {\phi}_{M^{\prime}}^{p}
  -{\phi}_{M^{\prime}}^{t} \big) \big]
   \nonumber \\ & & -\,
  {\phi}_{B2}\, x_{2}\, {\phi}_{M^{\prime}}^{a}\,
   \big({\phi}_{M}^{p}+{\phi}_{M}^{t}\big) \big\}_{b_{2}=b_{3}}
   \label{amp-a04-03},
   \end{eqnarray}
    \begin{equation}
    H_{cd}({\alpha},{\beta},b_{i},b_{j})\, =\,
    b_{i}\,b_{j}\, \big\{ {\theta}(b_{i}-b_{j})\,
    K_{0}\big(b_{i}\sqrt{{\alpha}}\big)\,
    I_{0}\big(b_{j}\sqrt{{\alpha}}\big)
    +\big(b_{i}\,{\leftrightarrow}\,b_{j}\big)
    \big\}\,K_{0}\big(b_{j}\sqrt{{\beta}}\big)
    \label{amp-e03-04},
    \end{equation}
    \begin{eqnarray}
    H_{ef}({\omega},{\beta},b_{i},b_{j}) &=&
    -\frac{{\pi}^{2}}{4}\,b_{i}\,b_{j}\,
    \big\{ {\theta}(b_{i}-b_{j})\, \big[
    J_{0}\big(b_{i}\sqrt{{\beta}}\big)
    +i\,Y_{0}\big(b_{i}\sqrt{{\beta}}\big) \big]\,
    J_{0}\big(b_{j}\sqrt{{\beta}}\big)
    \nonumber \\ & & \hspace{0.1\textwidth}
    +\big(b_{i}\,{\leftrightarrow}\,b_{j}\big) \big\}
    \big\{ J_{0}\big(b_{i}\sqrt{\omega}\big)
    +i\,Y_{0}\big(b_{i}\sqrt{\omega}\big) \big\}
    \label{amp-a01-04},
    \end{eqnarray}
    \begin{eqnarray}
    H_{gh}({\omega},{\beta},b_{i},b_{j}) &=&
    i\,\frac{{\pi}}{2}\,b_{i}\,b_{j}\, \big\{ \big[
     J_{0}\big(b_{i}\sqrt{{\omega}}\big)+i\,
     Y_{0}\big(b_{i}\sqrt{{\omega}}\big) \big]\,
     J_{0}\big(b_{j}\sqrt{{\omega}}\big)
    +\big(b_{i}\,{\leftrightarrow}\,b_{j}\big) \big\}
    \nonumber \\ & & \hspace{-0.01\textwidth}
    \big\{ i\,\frac{\pi}{2}\,{\theta} ({\beta})\,\big[
     J_{0}\big(b_{i}\sqrt{ {\beta}}\big)+i\,
     Y_{0}\big(b_{i}\sqrt{ {\beta}}\big) \big] +
   {\theta}(-{\beta})\,K_{0}\big(b_{i}\sqrt{-{\beta}}\big) \big\}
    \label{amp-a03-04},
    \end{eqnarray}
   \begin{equation}
  {\alpha}_{g}\, =\, x_{1}\,x_{2}\,m_{B}^{2}
   \label{amp-e01-05},
   \end{equation}
   \begin{equation}
  {\omega}_{g}\, =\, x_{2}\,\bar{x}_{3}\,m_{B}^{2}
   \label{amp-a01-05},
   \end{equation}
   \begin{equation}
  {\beta}_{a}\, =\, x_{2}\,m_{B}^{2}
   \label{amp-e01-06},
   \end{equation}
   \begin{equation}
  {\beta}_{b}\, =\, x_{1}\,m_{B}^{2}
   \label{amp-e02-06},
   \end{equation}
   \begin{equation}
  {\beta}_{c}\, =\,x_{2}\, ( x_{1}-\bar{x}_{3} )\, m_{B}^{2}
   \label{amp-e03-06},
   \end{equation}
   \begin{equation}
  {\beta}_{d}\, =\,x_{2}\,( x_{1}-x_{3} )\, m_{B}^{2}
   \label{amp-e04-06},
   \end{equation}
   \begin{equation}
  {\beta}_{e}\, =\, \bar{x}_{3}\,m_{B}^{2}
   \label{amp-a01-06},
   \end{equation}
   \begin{equation}
  {\beta}_{f}\, =\, x_{2}\,m_{B}^{2}
   \label{amp-a02-06},
   \end{equation}
   \begin{equation}
  {\beta}_{g}\, =\, ( x_{2}\,\bar{x}_{3}-\bar{x}_{1}\,x_{2}
  -\bar{x}_{3} )\,m_{B}^{2}
   \label{amp-a03-06},
   \end{equation}
   \begin{equation}
  {\beta}_{h}\, =\, x_{2}\,( \bar{x}_{3}-x_{1} )\,m_{B}^{2}
   \label{amp-a04-06},
   \end{equation}
   \begin{equation}
   t_{i}= {\max}\big(1/b_{1},1/b_{2},\sqrt{{\beta}_{i}}\big),
   \ \text{for}\ i=a,b
   \label{amp-e01-07};
   \end{equation}
   \begin{equation}
   t_{i}= {\max}\big(1/b_{2},1/b_{3},\sqrt{{\alpha}_{g}},
        \sqrt{{\vert}{\beta}_{i}{\vert}}\big),\ \text{for}\ i=c,d
   \label{amp-e03-07};
   \end{equation}
   \begin{equation}
   t_{i}= {\max}\big(1/b_{2},1/b_{3},\sqrt{{\beta}_{i}}\big),
   \ \text{for}\ i=e,f
   \label{amp-a01-07};
   \end{equation}
   \begin{equation}
   t_{i}= {\max}\big(1/b_{1},1/b_{2},\sqrt{{\omega}_{g}},
        \sqrt{{\vert}{\beta}_{i}{\vert}}\big),\ \text{for}\ i=g,h
   \label{amp-a03-07}.
   \end{equation}
  \end{appendix}

  
  \end{document}